\documentclass[12pt]{article}
\usepackage{amsfonts}
\usepackage{amsmath,amsthm,amssymb}
\usepackage{geometry}
\usepackage{graphicx}
\usepackage{subfigure}
\usepackage{hyperref}
\usepackage{cancel}
\usepackage{textcomp}
\usepackage{tikz}
\usepackage{bm}
\usepackage{mathrsfs}
\usepackage{filecontents}
\usepackage{times}
\usepackage{epsfig}
\usepackage{xcolor}
\usepackage{slashed}
\usepackage{booktabs}
\usepackage{latexsym}
\usepackage{verbatim}
\usepackage{extarrows}
\usepackage{multirow}
\usepackage{rotating}
\usepackage{colortbl}
\usepackage{indentfirst}
\usepackage[numbers,sort&compress]{natbib}
\usepackage{soul}
\definecolor{mygray}{gray}{.9}
\definecolor{intnull}{RGB}{213,229,255}
\usepackage{arydshln}
\usepackage{diagbox}
\usepackage{appendix}

\newcommand{\dif}{\mathrm{d}}

\usepackage{caption}
\usepackage{tikz}
\usetikzlibrary{arrows,shapes,chains}
\begin{document}
\renewcommand{\thefootnote}{\fnsymbol{footnote}}
\baselineskip=16pt
\pagenumbering{arabic}
\vspace{1.0cm}
\begin{center}
{\Large\sf Echoes and quasi-normal modes of perturbations around Schwarzschild traversable wormholes}
\\[10pt]
\vspace{.5 cm}
{{Hao Yang${}^{1,2,}$\footnote{E-mail address: hyang@ucas.ac.cn}}, 
{Zhong-Wu Xia${}^{2,}$\footnote{E-mail address: 2120210160@mail.nankai.edu.cn}},
and
{Yan-Gang Miao${}^{2,3,}$\footnote{Corresponding author. E-mail address: miaoyg@nankai.edu.cn}}

\vspace{6mm}
${}^{1}${\normalsize \em
School of Fundamental Physics and Mathematical Sciences, Hangzhou Institute for Advanced Study, UCAS, Hangzhou 310024, China}

${}^{2}${\normalsize \em School of Physics, Nankai University, Weijin Road 94, Tianjin 300071, China}

\vspace{3mm}
${}^{3}${\normalsize \em Faculty of Physics, University of Vienna, Boltzmanngasse 5, A-1090 Vienna, Austria}
}

\vspace{4.0ex}
\end{center}
\begin{center}
{\bf Abstract}
\end{center}

We investigate the waveforms and quasi-normal modes around Schwarzschild traversable wormholes under different field perturbations, including the scalar field, the electromagnetic (vector) field, and the axial gravitational (tensor) field perturbations.
Our results indicate that under the influence of the matter at the throat of the wormhole, a Dirac $\delta$-function distribution of matter appears in the effective potentials of the scalar and axial gravitational perturbations and it affects the propagation of these two types of perturbations in spacetime.
However, the matter at the throat has no influence on the propagation of electromagnetic perturbations.
Furthermore, we quantify the impact of throat matter on both the perturbation waveforms and quasi-normal modes for all three field types.
Through comparative studies between Schwarzschild traversable wormholes and Schwarzschild black holes, we identify two distinct features.
Firstly, the perturbation waveforms exhibit echoes and damping oscillations around wormholes, whereas they solely display damping oscillations around black holes.
Secondly, the difference between the adjacent peaks varies with the mass parameter and the throat radial coordinate in the waveform around Schwarzschild traversable wormholes, while  a constant peak spacing occurs, which is determined solely by mass, in the waveform around Schwarzschild black holes.
Based on these findings, we propose a framework to estimate the mass parameter and throat radial coordinate of Schwarzschild traversable wormholes through waveforms and quasi-normal modes.
Our analyses provide a more profound comprehension of the inherent characteristics of Schwarzschild traversable wormholes.

\renewcommand{\thefootnote}{\arabic{footnote}}
\newpage
\tableofcontents

\newpage
\section{Introduction}
\label{sec:intrN}
A wormhole is a spacetime structure connecting two regions in our universe or a multiverse.
The concept of wormholes dates back nearly a century, with early theoretical models describing untraversable wormholes~\cite{weyl1949philosophy,einstein1935particle} due to instabilities, tidal gravitational forces, and other limitations.
In the 1980s, traversable wormhole solutions were proposed~\cite{Morris:1988cz,Morris:1988tu}, marking a significant theoretical advance.
Shortly thereafter, a general method to construct traversable wormholes, called the ``cut-and-paste" technique, was introduced~\cite{Visser:1989kg,Visser:1989kh,Visser:1990wi,Konoplya:2021hsm,Lobo:2020ffi}. 
The core idea of this method is to join two identical template spacetime manifolds along a shared hypersurface.
To date, this method has been widely adopted~\cite{Perry:1991qq,Bhawal:1992sz,Richarte:2009zz,Eiroa:2009hm,Eiroa:2008hv,Lemos:2008aj,Rahaman:2006xb,Rahaman:2006vg,Thibeault:2005ha,Bejarano:2011yz,Rahaman:2010bt,Magalhaes:2023har,Javed:2025wty,Mazharimousavi:2025vvj} for building traversable wormholes within various theoretical frameworks.
Wormholes constructed via this technique typically exhibit a Dirac $\delta$-function distribution of exotic matter at the throat, earning them the designation ``thin-shell wormholes". 
However, traversable wormholes remain purely theoretical constructs and no experimental evidence has yet confirmed their existence.	

At present, it has gradually become feasible to verify gravitational theories from observations after the continuous development of gravitational wave detection technology~\cite{LIGOScientific:2016aoc,LIGOScientific:2017vwq,LIGOScientific:2021qlt}.
Moreover, the difference of spacetime structures between a wormhole (constructed by the ``cut-and-paste" technique) and a template spacetime is also expected~\cite{Chen:2022tog,Guo:2022iiy,Bao:2022iaz,GalvezGhersi:2019lag,DeFalco:2023kqy,DeFalco:2021klh,Bronnikov:2021liv,Dai:2019mse,Churilova:2021tgn,Konoplya:2016hmd,Bronnikov:2012ch} to be distinguished by observations.
In the present work, we start with a Schwarzschild traversable wormhole derived~\cite{Visser:1989kg} from a Schwarzschild spacetime as the template.
Then we try to investigate the difference between a Schwarzschild traversable wormhole and a Schwarzschild black hole through two key observables: quasi-normal modes (QNMs) and echoes.

QNMs are critical observables used to describe the waveform of perturbations in spacetime and differentiate compact objects.
Within the framework of spacetime perturbation theory~\cite{Chandrasekhar:1975zza,Berti:2009kk,Blazquez-Salcedo:2018ipc,Azad:2022qqn}, a perturbed spacetime generates characteristic waveforms that are directly linked to its dynamical stability.
In the systems approaching stability, these waveforms exhibit damping oscillatory behavior – periodic oscillations with exponentially decaying amplitudes.
Such a damping oscillation perturbation waveform is usually called a quasi-normal mode whose complex eigenfrequencies are expressed as
\begin{equation}\label{QNM}
\omega=\omega_{\rm R}+i\omega_{\rm I},
\end{equation}
where $\omega_{\rm R}$ denotes the oscillation frequency and $\omega_{\rm I}$ the decay rate. 
Eigenfrequencies of QNMs encode intrinsic spacetime properties (whether black holes or traversable wormholes) and are uniquely determined by metric parameters. 
Consequently, the QNM analysis remains an important issue in black hole physics~\cite{Cartwright:2024iwc,Wang:2024gzr,Chung:2024vaf,Livine:2024bvo}. 
Meanwhile, the comparative analyses of QNMs between wormholes and black holes continue to emerge~\cite{DeSimone:2025sgu,Cardoso:2016rao,Damour:2007ap,Konoplya:2016hmd,Konoplya:2005et}.
When a spacetime structure changes from a black hole to a traversable wormhole, its corresponding QNMs alter~\cite{Cardoso:2016rao} accordingly, providing us the possibility to differentiate these two compact objects.

Furthermore, the perturbation waveforms in wormhole spacetimes may exhibit echo phenomena~\cite{Liu:2020qia,DuttaRoy:2019hij,Ou:2021efv,Yang:2021cvh,Biswas:2022wah}. 
These waveforms are characterized by periodic pulse-shaped enhancements, which are absent in Schwarzschild black hole backgrounds.
The emergence of echoes stems from the multi-barrier structure of perturbation's effective potentials, which is fundamentally governed by both spacetime geometry and matter distribution. 
This unique correlation makes echo analyses particularly valuable for spacetime characterization, hence its prominence in studies of exotic compact objects~\cite{Rosato:2025byu,Shen:2024rup,Qian:2024zvq,Biswas:2023ofz,Lin:2023qgd}.
By systematically comparing perturbation waveforms and QNMs between Schwarzschild traversable wormholes and Schwarzschild black holes, we establish diagnostic criteria for identifying wormhole signatures. 
Additionally, we investigate how throat matter configurations influence perturbation fields, quantifying their effects on both waveforms and QNM spectra.

The paper is organized as follows.
In Sec.~\ref{sec:Schwarzschild sol}, we make a detailed analysis of Schwarzschild traversable wormholes.
In Sec.~\ref{sec:PEFDM}, we give perturbation equations under different field perturbations and introduce the finite difference method.
In Sec.~\ref{sec:scalar}, we analyze the waveforms and QNMs under the scalar field perturbations and propose a scenario to estimate the parameters of Schwarzschild traversable wormholes.
In Sec.~\ref{sec:EM} and Sec.~\ref{sec:Gravity}, we further investigate the waveforms and QNMs for Schwarzschild traversable wormholes under the electromagnetic field perturbations and the axial gravitational field perturbations, respectively. 
Finally, we give our conclusion in Sec.~\ref{sec:Con}. In addition, two appendices are given, one is the derivation of scalar, electromagnetic, and axial gravitational field perturbation equations in Appendix~\ref{appendix:A}, and the other is the derivation of difference equations in Appendix~\ref{appendix:B}.

\section{Schwarzschild traversable wormholes}\label{sec:Schwarzschild sol}
\subsection{Construction of Schwarzschild traversable wormholes}\label{sec:Costw}
Schwarzschild traversable wormholes are constructed~\cite{Visser:1989kg} by the ``cut-and-paste" technique together with the Schwarzschild spacetime serving as a template.
The main steps are briefly introduced as follows.
At first, we consider the Schwarzschild spacetime described by the metric, 
\begin{equation}\label{eq:metricSch}
   \dif s^2=-\left(1-\frac{2M}{\chi}\right)\dif t^2+ \left(1-\frac{2M}{\chi}\right)^{-1}\dif \chi^2+ \chi^2\left(\dif \theta^2 +\sin^2\theta\dif\phi^2\right),
\end{equation}
where $M$ is the mass, $\chi$ the radial coordinate, $\chi\in[0,+\infty)$,
and the horizon is located at $\chi=2M$.
Then, we take two copies of this manifold, labeled with ``$1$" and ``$2$", respectively, and dig out two four-dimensional regions, $\Omega_i\equiv\{\chi_i< b|b> 2M\}$, $i=1,2$, in the two copied manifolds, where $b$ is a constant and is taken to be $b>2M$  to ensure that there are no horizons in the remaining spacetimes.
Now each manifold is a geodesically incomplete manifold with a time-like hypersurface boundary, $\partial\Omega_i\equiv\{\chi_i=b|b> 2M\}$, $i=1,2$.
Finally, we combine these two manifolds by taking $\partial\Omega_1=\partial\Omega_2\equiv\partial\Omega$ and obtain a geodesically complete wormhole manifold with the throat at $\partial\Omega$.

The above process can be implemented by making the transformation $\chi\to\chi(r)$ in Eq.~\eqref{eq:metricSch} and obtaining the corresponding metric,
\begin{equation}\label{eq:metric_general}
\dif s^2=-\left[1-\frac{2M}{\chi(r)}\right]\dif t^2+\left[1-\frac{2M}{\chi(r)}\right]^{-1}\chi'^2(r)\dif r^2+\chi^2(r)(\dif\theta^2+\sin^2\theta\dif\phi^2),
\end{equation}
where $r$ is regarded as a parameter, $r\in(-\infty,+\infty)$, the prime indicates the derivative with respect to $r$, and $\chi(r)$ satisfies the following three conditions:
\begin{itemize}
    \item $\chi(r)$ is an even function to ensure that the metric for $r<0$ is a copy of that for $r>0$ and both are copies of a Schwarzschild spacetime.
    \item $\dif\chi(r)/\dif r$ is always greater than zero when $r > 0$, ensuring that $\chi(r)$ and $r$ have a one-to-one correspondence.
    \item $\chi(0)=a$, which means that we remove the region $\chi<a$ from a Schwarzschild spacetime and connect two identical spacetimes\footnote[1]{Here, the spacetime means the remaining part after a region of the Schwarzschild spacetime is removed by the four-dimensional domain, $\Omega\equiv\{\chi<a|a>0\}$.} at $r=0$. Here, $a$ is a parameter.
\end{itemize}

When we take $a = b > 2M$, we achieve the construction of the wormhole in the same way as the ``cut-and-paste" technique.
From now on, we no longer assume $a> 2M$ because we want to analyze how the event horizon affects the observations of wormholes.
The spacetime described by Eq.~(\ref{eq:metric_general}) has the following three structures depending on the value of $a$: 
\begin{itemize}
    \item When $0<a<2M$, there exists a horizon, $\chi=2M$, and Eq.~\eqref{eq:metric_general} describes a black hole.
    In this situation, the wormhole throat located at $\chi=a$ is hidden by the horizon.
    However, there is no distinction between this solution and a Schwarzschild black hole outside the event horizon.
    Since the region within the event horizon cannot be observed from outside, we are unable to determine whether the traversable wormhole exists or not.
   \item When $a=2M$, the coordinate speed of light vanishes at $r=0$, i.e., $\frac{\dif \chi(r)}{\dif t}\big|_{r=0}=0$, where we choose $\chi(r)$ as the radial coordinate since it represents the radius in spacetime.
   At this time, the solution is not a black hole, but rather a one-way wormhole with a null-like throat located at $\chi=2M$.
   However, considering the similarity between the null-like throat and the event horizon, we are still unable to determine whether a traversable wormhole exists or not.
   \item When $a>2M$, there are no horizons and $\frac{\dif \chi(r)}{\dif t}\neq0$ in the region of $r\in(-\infty,+\infty)$. 
   Eq.~\eqref{eq:metric_general} describes~\cite{visser1995lorentzian} a traversable wormhole whose throat is located at $\chi(0)=a$, where the negative values of $r$ represent a geometric universe on the opposite side of the observer’s own universe.
   In this situation, the ``cut-and-paste" technique works, which will be confirmed by distinguishing a traversable wormhole from a black hole.
   Such changes in spacetime structures will be analyzed through the alterations of observables, e.g., echoes and QNMs. 
\end{itemize}

\subsection{Source of Schwarzschild traversable wormholes}\label{sec:Soctw}
Now we discuss the matter source of Schwarzschild traversable wormholes. 
If the first-order derivative of $\chi(r)$ with respect to $r$ is continuous at the throat of the wormhole, then $\dif\chi(r)/\dif r$ must be zero at the throat since $\chi(r)$ is an even function.
According to the metric Eq.~\eqref{eq:metric_general}, $g_{rr}=0$ at the throat of the wormhole, which makes the curvature term at the throat difficult to calculate. 
Therefore, we need to perform the calculation in an appropriate coordinate system.
Here, we adopt the tortoise coordinate $r_*$ defined by
\begin{align}\label{wh-tc}
    & \frac{\dif \chi}{\dif r_*}\equiv\left\{
    \begin{aligned}
        & 1-\frac{2M}{\chi},  &\text{at the observer's universe},\\
        & -\left(1-\frac{2M}{\chi}\right),  &\text{at the opposite universe}.
    \end{aligned}
    \right .
\end{align}
There are two motivations for adopting this coordinate system. 
One is that it facilitates the calculation of the curvature terms of spacetime. 
The other is that it is consistent with the coordinate used in the perturbation equation, which is conducive to the derivation of perturbation equations.
It can be seen that $\dif\chi/\dif r_*$ is not continuous at the throat of the wormhole, which will lead to the second-order derivative of $\chi$ with respect to $r_*$ taking the following form,
\begin{equation}\label{chi''}
    \frac{\dif^2\chi}{\dif r_*^2}=\left\{
    \begin{aligned}
        & \frac{2(a-2M)}{a}\delta(0),  &\text{at the throat},\\
        & \frac{2M}{\chi^3}(\chi-2M),  &\text{outside the throat},
    \end{aligned}
    \right .
\end{equation}
where $\delta(r_*)$ is the Dirac $\delta$-function.
In the tortoise coordinate, the metric of Schwarzschild traversable wormholes is
\begin{equation}\label{tc-metric}
    \dif s^2=-\left[1-\frac{2M}{\chi(r_*)}\right]\dif t^2+\left[1-\frac{2M}{\chi(r_*)}\right]\dif r_*^2+\chi^2(r_*)(\dif\theta^2+\sin^2\theta\dif\phi^2).
\end{equation}
And the nonzero components of the corresponding Ricci curvature tensor are as follows:
\begin{subequations}\label{rmunu}
\begin{equation}\label{r00}
    R_{00}=-\frac{2M^2}{\chi^4}+\frac{M}{\chi(\chi-2M)}\chi'',
\end{equation}
\begin{equation}\label{r11}
    R_{11}=\frac{2M^2}{\chi^4}+\frac{4M}{\chi^4}(\chi-2M)-\frac{2}{\chi}\chi''-\frac{M}{\chi(\chi-2M)}\chi'',
\end{equation}
\begin{equation}\label{r2233}
    R_{22}=\frac{1}{\sin^2\theta}R_{33}=\frac{2M}{\chi}-\frac{\chi^2}{\chi-2M}\chi'',
\end{equation}
\end{subequations}
and the corresponding Ricci scalar is
\begin{equation}
    R=g^{\mu\nu}R_{\mu\nu}=2R^1_1=\frac{4M^2}{\chi^3(\chi-2M)}+\frac{8M}{\chi^3}-\frac{2M}{(\chi-2M)^2}\chi''-\frac{4}{\chi-2M}\chi'',
\end{equation}
where $\chi''=\dif^2\chi/\dif r_*^2$.
In terms of the Einstein field equations,
\begin{equation}\label{Einstein-eq}
R_{\mu\nu}-\frac{1}{2}g_{\mu\nu}R=8\pi T_{\mu\nu},
\end{equation}
we can confirm that the matter source exists only at the throat $\chi=a$ in Schwarzschild traversable wormholes, and the nonzero components of the corresponding energy-momentum tensor take the following form,
\begin{subequations}\label{Tmunu}
\begin{equation}\label{T00}
    T_{00}(\chi=a)=\frac{M(a-2M)}{2\pi a^4}-\frac{a-2M}{2\pi a^2}\delta(0),
\end{equation}
\begin{equation}
    T_{22}(\chi=a)=\frac{1}{\sin^2\theta}T_{33}(\chi=a)=-\frac{M(a-M)}{4\pi a(a-2M)}+\frac{a(a-M)}{4\pi(a-2M)}\delta(0).
\end{equation}
\end{subequations}
After applying the appropriate coordinate transformation and neglecting the finite terms, we can verify that this energy-momentum tensor is consistent with the result obtained in Ref.~\cite{Visser:1989kg}.

The matter in the throat of a Schwarzschild traversable wormhole is~\cite{Visser:1989kg} essentially a shell of exotic matter with a concentrated negative energy density, which maintains the stability of the wormhole by violating the weak energy conditions.
The physical interpretation of this exotic matter is currently not unique. 
The proposed explanations encompass fluid models based on various equations of state (EoS)~\cite{Varela:2013xua}, quintessence-like matter~\cite{Javed:2025wty,Sharif:2021sjm}, and others~\cite{Thibeault:2005ha,Richarte:2009zz}.

\section{Perturbation equation and finite difference method}\label{sec:PEFDM}
\subsection{Perturbation equation}\label{sec:PE}
Generally, three types of fields are used to perturb~\cite{Berti:2009kk} a spacetime, which are the scalar field perturbation, the electromagnetic (vector) field perturbation, and the axial gravitational (tensor) field perturbation.
The wave function $\Phi_s$ of the three perturbation fields satisfies~\cite{Konoplya:2011qq,Berti:2009kk} the following Schr\"odinger-like equation,
\begin{equation}\label{eq:Peq}
\frac{\dif^2\Phi_s}{\dif r_*^2}-\frac{\dif^2\Phi_s}{\dif t^2}-V_s\Phi_s=0,
\end{equation}
where $V_s$ stands for the effective potential, $s$ the spin with $s=0,1,2$ corresponding to scalar, vector, and tensor perturbation fields, respectively,
and $r_*$ the tortoise coordinate.
Then we solve the tortoise coordinate for black holes and traversable wormholes, respectively.
\begin{itemize}
\item 
For a black hole with horizons, the tortoise coordinate is defined by
\begin{equation}\label{bh-tc}
    \dif r_*\equiv \frac{\chi}{\chi-2M}\dif\chi.
\end{equation}
Solving this equation, we obtain
\begin{equation}\label{r*BH}
    r_*=\chi+2M \ln\left(\frac{\chi}{2M}-1\right),
\end{equation}
where its lower boundary is located at the outer horizon, $\chi=2M$, and its upper boundary is located at infinity, $\chi=+\infty$, i.e., the tortoise coordinate takes the range, $r_*\in(-\infty,+\infty)$. 
In other words, the boundaries of the equations of motion are determined by the outer event horizon and the infinity of the observer's universe.
\item Solving Eq.~\eqref{wh-tc} for a Schwarzschild traversable wormhole without horizons, we obtain
\begin{align}\label{r*WH}
    & r_*=\left\{
    \begin{aligned}
        & 2a+4M\ln\left(\frac{a}{2M}-1\right)-\chi-2M \ln\left(\frac{\chi}{2M}-1\right),& \text{at the observer's universe},\\
        & \chi+2M \ln\left(\frac{\chi}{2M}-1\right),& \text{at the opposite universe},
    \end{aligned}
    \right .
\end{align}
where the range  of $r_*$ is still $(-\infty,+\infty)$, but the wormhole's lower bound is at $\chi=+\infty$ in the observer's universe and upper bound is at $\chi(r)=+\infty$ in the opposite universe. 
This shows that the boundaries of equations of motion are determined by the infinity of the observer's own universe and the infinity of the observer's opposite universe, where the two universes are connected through the throat of the traversable wormhole.
And the throat of the traversable wormhole is located at $r_*=a+2M\ln\left(\frac{a}{2M}-1\right)$.
\end{itemize}

For a Schwarzschild black hole, the effective potential of scalar field perturbations is given~\cite{Berti:2009kk} by
\begin{equation}\label{eq_scalar}
V_{s=0}=\left(1-\frac{2M}{\chi(r_*)}\right)\left[\frac{l(l+1)}{\chi^2(r_*)}+\frac{2M}{\chi^3(r_*)}\right],
\end{equation}
where $l$ is the azimuthal number and it satisfies $l\geq s$.
The effective potential of electromagnetic field perturbations takes~\cite{Berti:2009kk} the form,
\begin{equation}\label{eq_EM}
V_{s=1}=\left(1-\frac{2M}{\chi(r_*)}\right)\frac{l(l+1)}{\chi^2(r_*)}.
\end{equation}
Gravitational field perturbations are divided into axial (odd-parity) and polar (even-parity) ones. 
The parity equals $(-1)^{l+1}$ for the former and $(-1)^l$ for the latter under the transformation of $\theta\to-\theta$, respectively.
The wave function of axial gravitational field perturbations $\Phi^-_{s=2}$, called the Regge-Wheeler function~\cite{regge1957stability}, has its corresponding effective potential,
\begin{equation}\label{eq_G_odd}
V^-_{s=2}=\left(1-\frac{2M}{\chi(r_*)}\right)\left[\frac{l(l+1)}{\chi^2(r_*)}-\frac{6M}{\chi^3(r_*)}\right].
\end{equation}

For a Schwarzschild traversable wormhole outside the throat, the effective potential forms of these three field perturbations are the same as those in the above-mentioned black hole case.
And the effective potential of electromagnetic field perturbations in the throat is the same as that in Eq.~\eqref{eq_EM}.
However, at the throat $\chi=a$, the effective potential of scalar field perturbations takes the form,
\begin{equation}
    \label{SP-throat}
    V_{s=0}\left(\chi=a\right)=\left(1-\frac{2M}{a}\right)\left[\frac{2}{a}\delta(0)+\frac{l(l+1)}{a^2}\right],
\end{equation}
and the effective potential of axial gravitational field perturbations is given by
\begin{equation}
\label{ogP-throat}
  V^-_{s=2}(\chi=a)=\frac{12M^2-8Ma}{a^4}+\frac{a-2M}{a^3}l(l+1)+\frac{2a-2M}{a^2}\delta(0).
\end{equation}
Therefore, the propagation of scalar field perturbations and that of axial gravitational field perturbations in spacetime will be affected by the matter at the throat of wormholes, but the propagation of electromagnetic field perturbations will not be influenced.

For the sake of generality, we adopt dimensionless quantities defined as follows:
\begin{equation}\label{eq:dimensionless}
     \bar{\chi}\equiv \frac{\chi}{2M}, \qquad \bar{t}\equiv \frac{t}{2M},\qquad \bar{a}\equiv \frac{a}{2M}, \qquad \bar{V}_s\equiv 4M^2V_s,\qquad \bar{r}_*\equiv \frac{r_*}{2M}.
\end{equation}
After using the above dimensionless quantities, we rewrite Eqs.~\eqref{eq:Peq},  (\ref{r*BH}), (\ref{r*WH}), and (\ref{eq_scalar})-(\ref{ogP-throat}) as follows:
\begin{equation}\label{eq:Peg-dl}
    \frac{\dif^2\Phi_s}{\dif \bar{r}_*^2}-\frac{\dif^2\Phi_s}{\dif \bar{t}^2}-\bar{V}_s\Phi_s=0,
\end{equation}
where the tortoise coordinate takes the forms,
\begin{equation}\label{r*BH-dl}
	\bar{r}_*=\bar{\chi}+\ln\left(\bar{\chi}-1\right),
\end{equation}
and
\begin{align}\label{r*WH-dl}
	& \bar{r}_*=\left\{
	\begin{aligned}
		& 2\bar{a}+2\ln\left(\bar{a}-1\right)-\bar{\chi}-\ln\left(\bar{\chi}-1\right),& \text{at the observer's universe},\\
		& \bar{\chi}+ \ln\left(\bar{\chi}-1\right),& \text{at the opposite universe},
	\end{aligned}
	\right.
\end{align}
for black holes and traversable wormholes, respectively. 
The effective potential of electromagnetic field perturbations reads
\begin{equation}\label{eq:EM-dl}
    \bar{V}_{s=1}=\left(1-\frac{1}{\bar{\chi}(\bar{r}_*)}\right)\frac{l(l+1)}{\bar{\chi}^2(\bar{r}_*)},
\end{equation}
which holds true for both wormholes and black holes.
For a black hole or a Schwarzschild traversable wormhole outside its throat, the effective potential of scalar field perturbations is 
\begin{equation}
    \bar{V}_{s=0}(\bar{\chi}\neq\bar{a})=\left(1-\frac{1}{\bar{\chi}(\bar{r}_*)}\right)\left[\frac{l(l+1)}{\bar{\chi}^2(\bar{r}_*)}+\frac{1}{\bar{\chi}^3(\bar{r}_*)}\right],\label{eq:scalar-ot-dl}
\end{equation}
and that of axial gravitational field perturbations is
\begin{equation}
    \bar{V}^-_{s=2}(\bar{\chi}\neq\bar{a})=\left(1-\frac{1}{\bar{\chi}(\bar{r}_*)}\right)\left[\frac{l(l+1)}{\bar{\chi}^2(\bar{r}_*)}-\frac{3}{\bar{\chi}^3(\bar{r}_*)}\right].\label{eq:oddgravity-ot-dl}
\end{equation}
And at the throat of wormholes, the effective potential of scalar field perturbations becomes
\begin{equation}
    \bar{V}_{s=0}(\bar{\chi}=\bar{a})=\left(1-\frac{1}{\bar{a}}\right)\left[\frac{l(l+1)}{\bar{a}^2}+\frac{2}{\bar{a}}\delta(0)\right],\label{eq:scalar-at-dl}
\end{equation}
and that of axial gravitational field perturbations changes to
\begin{equation}
    \bar{V}^-_{s=2}(\bar{\chi}=\bar{a})=\frac{3-4\bar{a}}{\bar{a}^4}+\frac{\bar{a}-1}{\bar{a}^3}l(l+1)+\frac{2\bar{a}-1}{\bar{a}^2}\delta(0).\label{eq:oddgravity-at-dl}
\end{equation}
We note that Eqs.~(\ref{eq:Peg-dl})-(\ref{eq:oddgravity-at-dl}) do not explicitly contain $M$, which means that the dimensionless treatment changes a double-parameter issue related to $M$ and $a$ into a single-parameter one related only to $\bar{a}$. 
This feature will greatly simplify our subsequent calculations and yield the results with generality.

\begin{figure}[htbp]
	\centering
	\subfigure[3D embedding diagram for $\bar{a}=1.1$]{
		\begin{minipage}[t]{0.4\linewidth}\label{Fig:3D-Embed}
			\centering
			\includegraphics[width=1\linewidth]{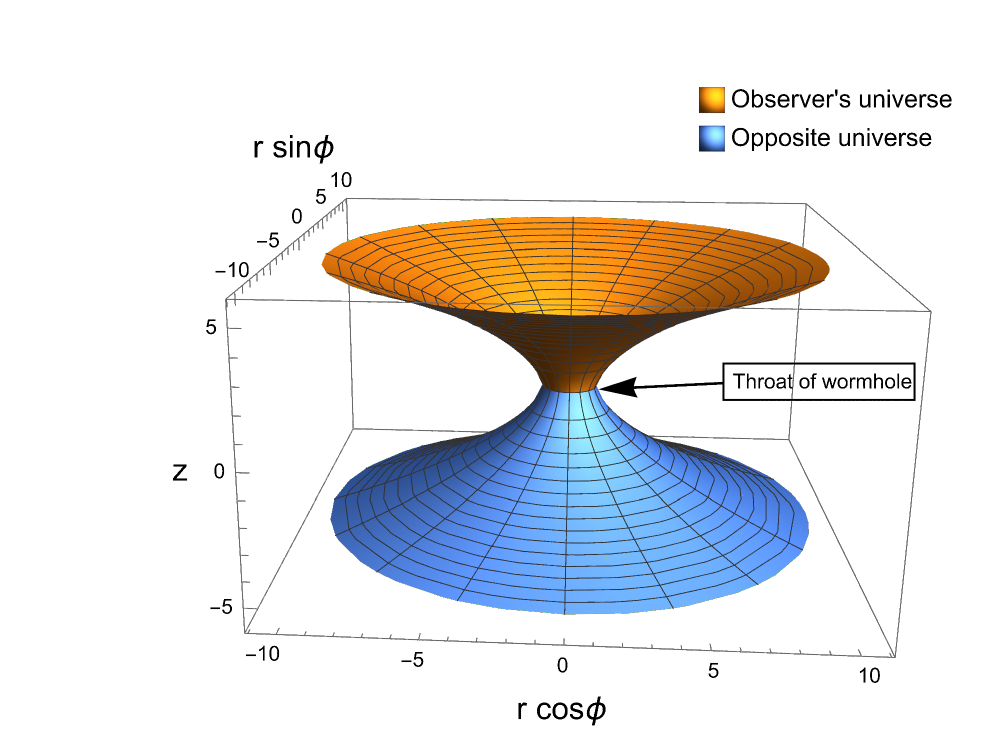}
		\end{minipage}
	}
	\subfigure[3D embedding diagram for $\bar{a}=1.5$]{
		\begin{minipage}[t]{0.4\linewidth}\label{Fig:3D-Embed-2}
			\centering
			\includegraphics[width=1\linewidth]{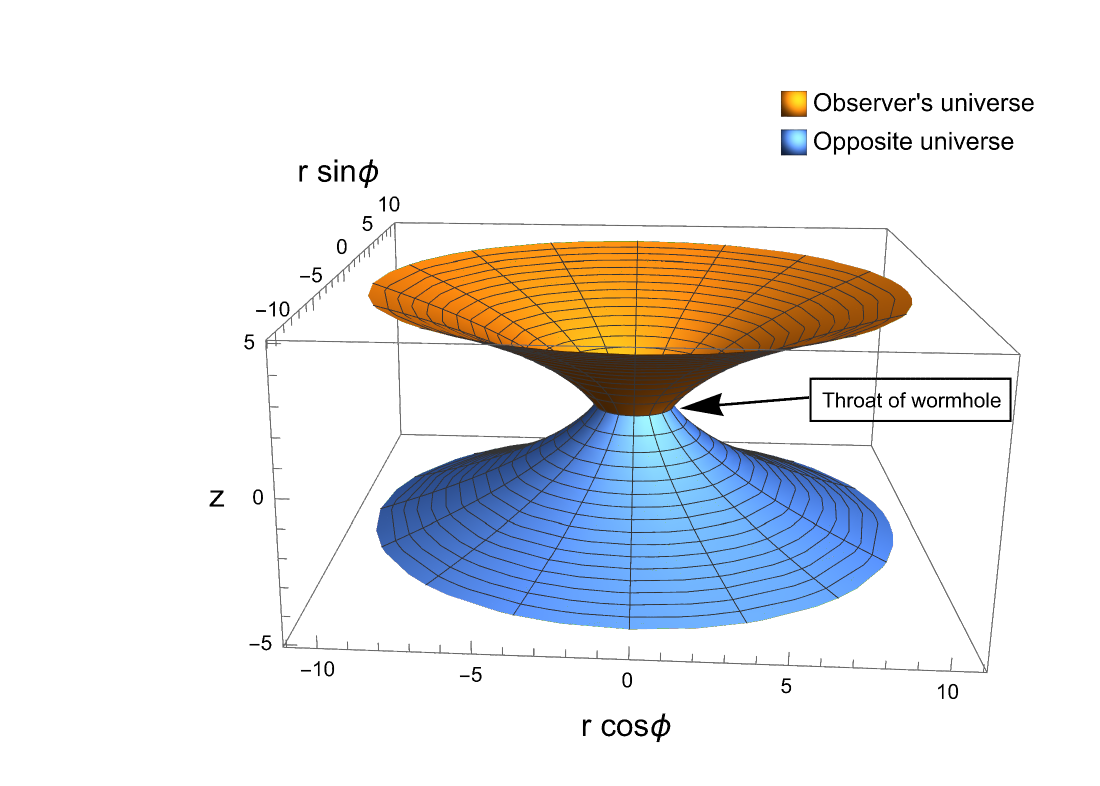}
		\end{minipage}
	}
	\vspace{-3mm}
	\subfigure[2D embedding diagram for $\bar{a}=1.1$]{
		\begin{minipage}[t]{0.4\linewidth}\label{Fig:2D-Embed}
			\centering
			\includegraphics[width=1\linewidth]{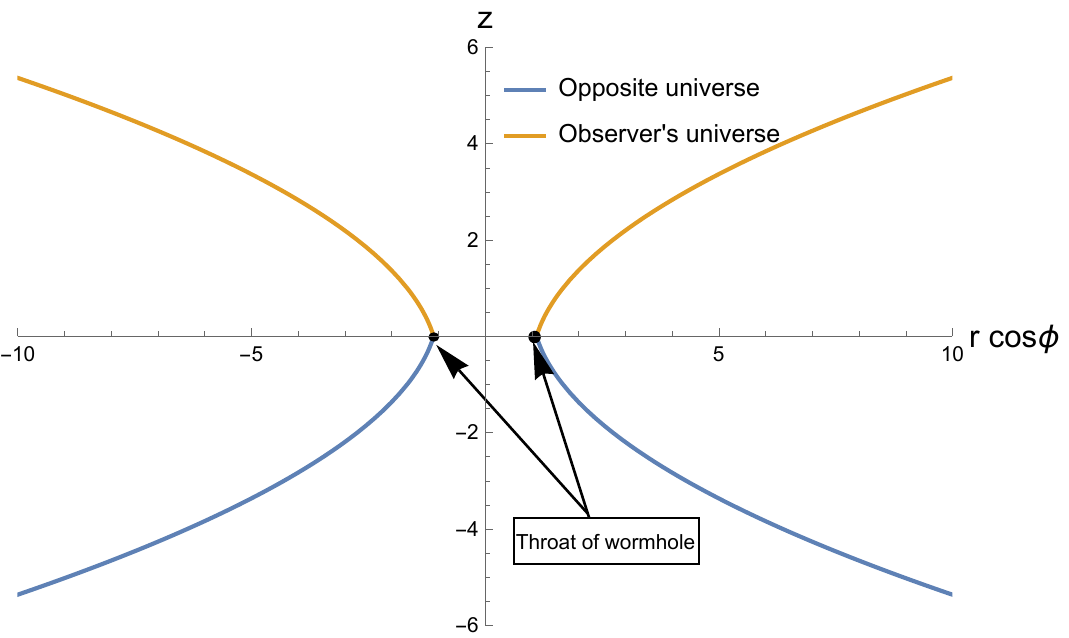}
		\end{minipage}
	}
\subfigure[2D embedding diagram for $\bar{a}=1.5$]{
		\begin{minipage}[t]{0.4\linewidth}\label{Fig:2D-Embed-2}
			\centering
			\includegraphics[width=1\linewidth]{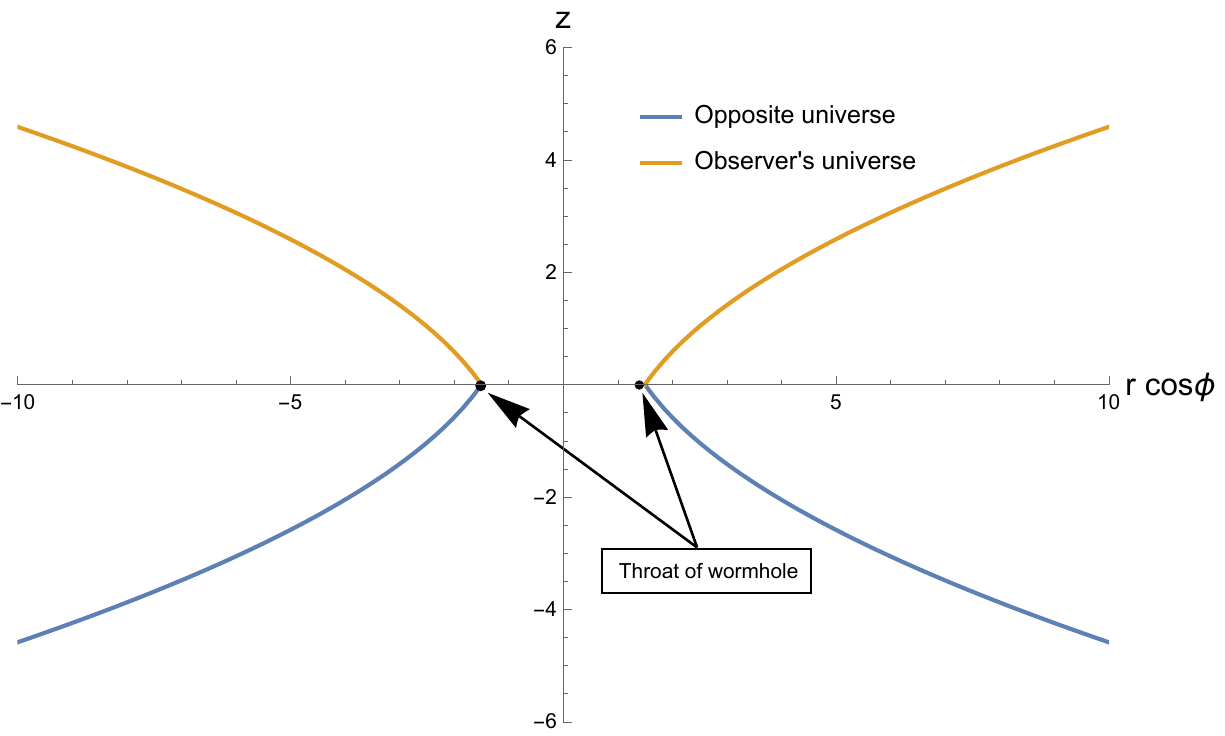}
		\end{minipage}
	}

\caption{
These  figures are the embedding diagrams of the Schwarzschild traversable wormholes with $\bar{a}=1.1, 1.5$. 
Among them, pictures (a) and (b) are the 3D embedding diagrams of the wormholes, which are drawn according to Eq.~\eqref{Eq:embed-surface}. 
The orange parts represent the observer's universes, and the blue ones represent the opposite universes. 
Pictures (c) and (d) are the 2D embedding diagrams of the wormholes, which are drawn by taking the cross section of pictures  (a) and (b) at $r\sin\phi=0$, respectively. 
The orange curves represent the observer's universes, and the blue ones represent the opposite universes.}
	\label{fig:embed}
\end{figure}

\begin{figure}[htbp]
	\centering
	\subfigure[$\bar{a}=1.1$]{
		\begin{minipage}[t]{0.4\linewidth}
			\centering\label{Fig:embed-matter}
			\includegraphics[width=1\linewidth]{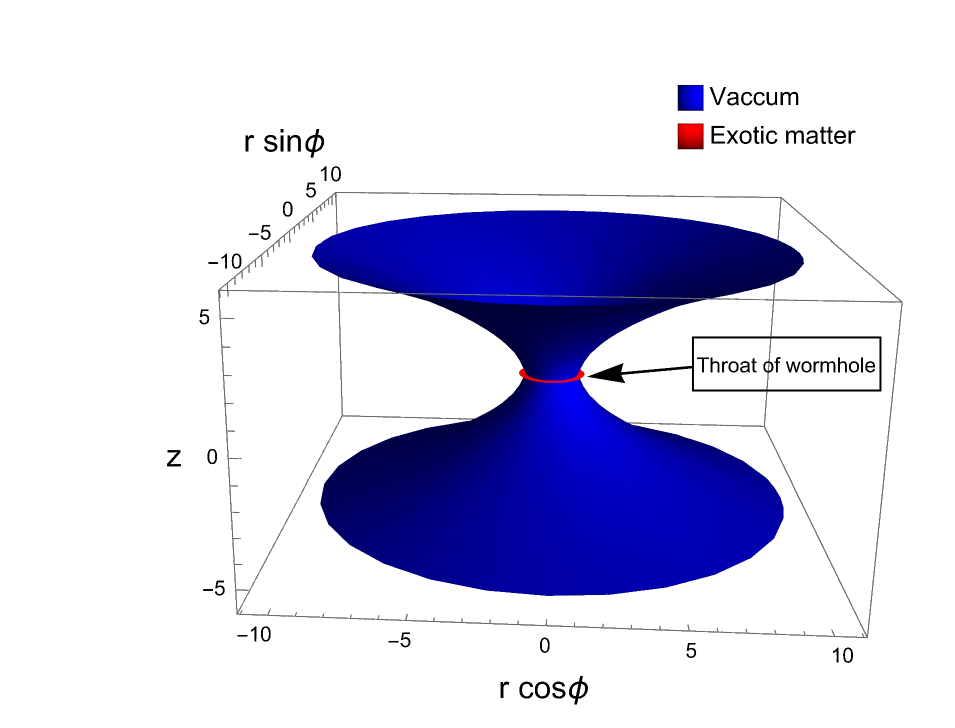}
		\end{minipage}
        }
    \subfigure[$\bar{a}=1.5$]{    
        \begin{minipage}[t]{0.4\linewidth}
			\centering\label{Fig:embed-matter-2}
			\includegraphics[width=1\linewidth]{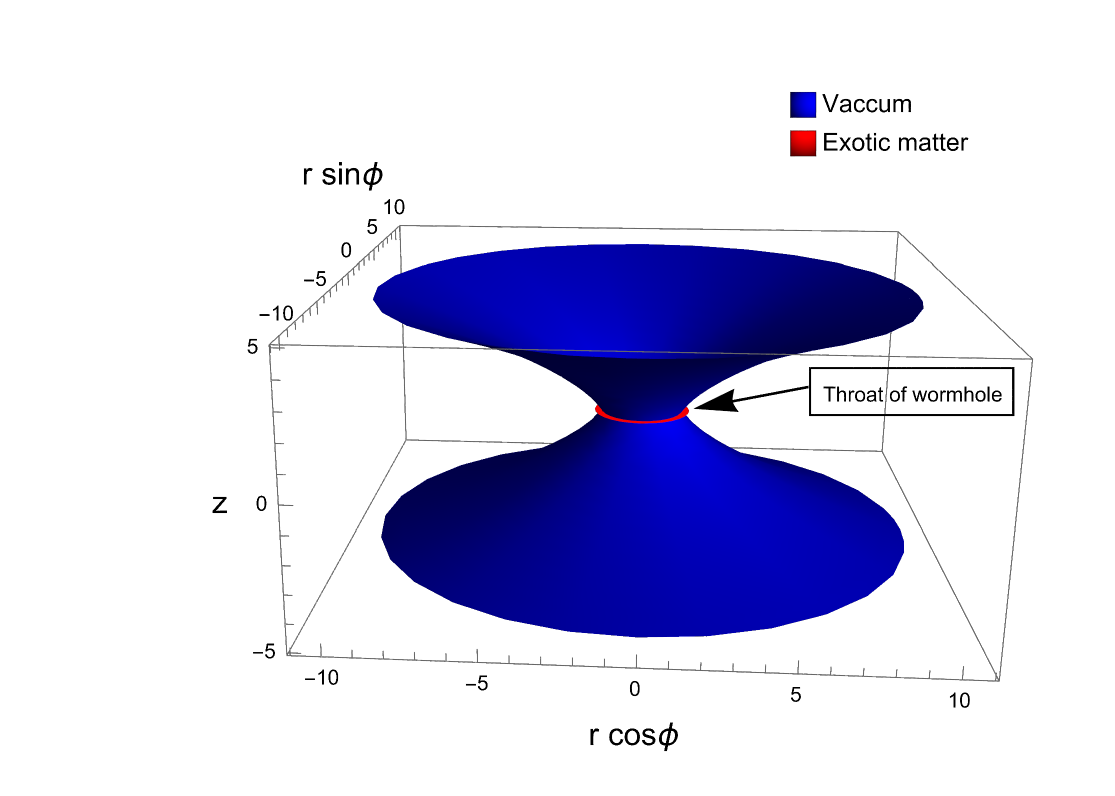}
		\end{minipage}
	}
	\caption{
These diagrams are schematic diagrams of the matter distributions of the Schwarzschild traversable wormholes with $\bar{a}=1.1, 1.5$. 
They are drawn based on the discussion of the wormhole matter source in Sec.~\ref{sec:Soctw}. 
The blue parts represent the vacuum, and the red ones, distributed on the throats of the wormholes, represent the exotic matter.}
\end{figure}

At the end of this subsection, we draw the 2D and 3D embedding diagrams of the wormhole in order to present the wormhole more clearly.
The 3D embedding diagram of a wormhole represents~\cite{Morris:1988cz}  the embedding of an equal-time and equal-$\theta$ slice of the wormhole into the three-dimensional Euclidean space.
Since Schwarzschild traversable wormholes are spherically symmetric, we can take $\theta = \pi/2$ without loss of generality.
By setting $t=\text{const.}$ and $\theta=\pi/2$ in Eq.~\eqref{eq:metricSch}, we give the dimensionless line element,
\begin{equation}\label{Eq:metric-slice}
   \dif \bar{s}^2=\left(1-\frac{1}{\bar{\chi}}\right)^{-1}\dif \bar{\chi}^2+ \bar{\chi}^2\dif \phi^2,
\end{equation}
where $\dif \bar{s}^2\equiv \frac{1}{4M^2}\dif s^2$, and then obtain its formulation
in the three-dimensional Euclidean space $(z, \bar{\chi}, \phi)$ as follows:
\begin{equation}\label{Eq:metric-Eu}
   \dif \bar{s}^2=\dif z^2+\dif \bar{\chi}^2+ \bar{\chi}^2 \dif \phi^2.
\end{equation}
On the embedded surface $z(\bar{\chi})$ of the wormhole in this Euclidean space, Eq.~\eqref{Eq:metric-Eu} degenerates into Eq.~\eqref{Eq:metric-slice}.
Therefore, the embedded surface $z(\bar{\chi})$ reads
\begin{equation}
    \label{Eq:embed-surface}
    z(\bar{\chi})=\pm 2\sqrt{\bar{\chi}-1}+C_{\pm},
\end{equation}
where $C_\pm$ are integral constants, and $\bar{\chi}\in[\bar{a},+\infty)$.
According to Eq.~\eqref{Eq:embed-surface}, we present the 3D embedding diagrams of the Schwarzschild traversable wormholes with $\bar{a}=1.1, 1.5$ in Figs.~\ref{Fig:3D-Embed} and \ref{Fig:3D-Embed-2}, respectively. 
Based on the two figures, we extract the cross sections where $r\sin\phi=0$ as their corresponding 2D embedding diagrams of the wormholes, as shown in Figs.~\ref{Fig:2D-Embed} and \ref{Fig:2D-Embed-2}.
Finally,  we plot the schematic diagrams of the matter distributions for the Schwarzschild traversable wormholes with $\bar{a}=1.1, 1.5$ in Figs.~\ref{Fig:embed-matter} and \ref{Fig:embed-matter-2}, respectively.

\subsection{Finite difference method}\label{sec:FDM}
When a spacetime structure changes, such as from a black hole to a traversable wormhole, the perturbation equation that associates with the spacetime structure will be varying accordingly, which inevitably alters the time-domain behavior of perturbations.
Here we employ the finite difference method~\cite{Ou:2021efv,Liu:2020qia} and the Prony method~\cite{Konoplya:2011qq,Berti:2007dg} to calculate perturbation waveforms and QNMs.

We start with discretizing the coordinate space $(\bar{t},\bar{r}_*)$, so that $(\bar{t},\bar{r}_*)$ becomes $(\bar{t}_0+j\Delta \bar{t},\bar{r}_{*0}+k\Delta \bar{r}_*)$, where $\bar{t}_0$ and $\bar{r}_{*0}$ are the coordinates of the grid origin, $j$ and $k$ are integers, and $\Delta \bar{t}$ and $\Delta \bar{r}_*$ are step sizes of coordinates.
For the sake of convenience, we write $(\bar{t}_0+j\Delta \bar{t},\bar{r}_{*0}+k\Delta \bar{r}_*)$ as $(j,k)$.
Meanwhile, we denote the values of $\bar{t}$ and $\bar{r}_*$ at the grid points $(j,k)$ as $\bar{t}_j$ and $\bar{r}_{*k}$.
When choosing the grid points, in order to avoid directly dealing with the $\delta$-function, we adjust $\bar{t}_0$ and $\bar{r}_{*0}$ so that the throat of the wormhole does not fall on the grid points.
If the grid does not pass through the throat of the wormhole, we can change the differential equation, Eq.~\eqref{eq:Peg-dl}, into the following difference equation as long as the step sizes are small enough,
\begin{eqnarray}\label{eq:Peg_FD}
& &
-\frac{\Phi_s(j+1,k)-2\Phi_s(j,k)+\Phi_s(j-1,k)}{\Delta \bar{t}^2}+\frac{\Phi_s(j,k+1)-2\Phi_s(j,k)+\Phi_s(j,k-1)}{\Delta \bar{r}_*^2}\nonumber \\
& &-\bar{V}_s(k)\Phi_s(j,k)=0,
\end{eqnarray}
from which we obtain the iterative formula of $\Phi_s$,
\begin{equation}\label{eq:iterative}
\Phi_s(j+1,k)=-\Phi_s(j-1,k)+\left[2-2\frac{\Delta \bar{t}^2}{\Delta \bar{r}_*^2}-\Delta \bar{t}^2\bar{V}_s(k)\right]\Phi_s(j,k)
+\frac{\Delta \bar{t}^2}{\Delta \bar{r}_*^2}\left[\Phi_s(j,k+1)+\Phi_s(j,k-1)\right].
\end{equation}

Then, we take $\bar{r}_t$ representing the position of the wormhole throat in the tortoise coordinate, that is, 
\begin{equation}
    \bar{r}_t=\bar{r}_*(\bar{\chi}=\bar{a})=\bar{a}+\ln(\bar{a}-1).
\end{equation}
If the grid does pass through the throat of the wormhole, i.e., $\bar{r}_t \in (\bar{r}_{*k-1}, \bar{r}_{*k})$, the difference equation of Eq.~\eqref{eq:Peg-dl} becomes\footnote[2]{The detailed derivation of the difference equation can be found in Appendix~\ref{appendix:B}.} 
\begin{eqnarray}\label{eq:Peg_FD-dat}
& &
-\frac{\Phi_s(j+1,k)-2\Phi_s(j,k)+\Phi_s(j-1,k)}{\Delta \bar{t}^2}+\frac{\Phi_s(j,k+1)-2\Phi_s(j,k)+\Phi_s(j,k-1)}{\Delta \bar{r}_*^2}\nonumber \\
& &-\bar{V}_s(k)\Phi_s(j,k)+\frac{C_s(\bar{a})(\bar{r}_{*k-1}-\bar{r}_t)\Phi_s(j,\bar{r}_t)}{\Delta \bar{r}^2_*}=0,
\end{eqnarray}
and the iterative formula is
\begin{equation}\label{eq:iterative-dpt}
\begin{split}
\Phi_s(j+1,k)=&-\Phi_s(j-1,k)+\left[2-2\frac{\Delta \bar{t}^2}{\Delta \bar{r}_*^2}-\Delta \bar{t}^2\bar{V}_s(k)\right]\Phi_s(j,k)\\
&+\frac{\Delta \bar{t}^2}{\Delta \bar{r}^2_*}C_s(\bar{a})(\bar{r}_{*k-1}-\bar{r}_t)\Phi_s(j,\bar{r}_t)
+\frac{\Delta \bar{t}^2}{\Delta \bar{r}_*^2}\left[\Phi_s(j,k+1)+\Phi_s(j,k-1)\right],
\end{split}
\end{equation}
where $C_s(\bar{a})$ is the coefficient of the $\delta$-function term in the effective potential at the throat,
\begin{equation}\label{csa}
    C_s(\bar{a})=\left\{
    \begin{aligned}
        & \frac{2(\bar{a}-1)}{\bar{a}^2},&s=0;\\
        & 0,&s=1 ;\\
        & \frac{2\bar{a}-1}{\bar{a}^2},&s=2,
    \end{aligned}
    \right .
\end{equation}
and $s=0,1,2$  represent scalar field perturbations, electromagnetic field perturbations, and axial gravitational field perturbations, respectively, and the factor in Eq.~\eqref{eq:iterative-dpt}, $\Phi_s(j,\bar{r}_t)$ satisfies
\begin{equation}
    \Phi_s(j,\bar{r}_t)=\frac{(\bar{r}_t-\bar{r}_{*k-1})\Phi_s(j,k)-(\bar{r}_t-\bar{r}_{*k})\Phi_s(j,k-1)}{\Delta \bar{r}_*-C_s(\bar{a})(\bar{r}_t-\bar{r}_{*k-1})(\bar{r}_t-\bar{r}_{*k})}.
\end{equation}
And when $\bar{r}_t \in (\bar{r}_{*k}, \bar{r}_{*k+1})$, the difference equation of Eq.~\eqref{eq:Peg-dl} is
\begin{eqnarray}\label{eq:Peg_FD-uat}
& &
-\frac{\Phi_s(j+1,k)-2\Phi_s(j,k)+\Phi_s(j-1,k)}{\Delta \bar{t}^2}+\frac{\Phi_s(j,k+1)-2\Phi_s(j,k)+\Phi_s(j,k-1)}{\Delta \bar{r}_*^2}\nonumber \\
& &-\bar{V}_s(k)\Phi_s(j,k)-\frac{C_s(\bar{a})(\bar{r}_{*k+1}-\bar{r}_t)\Phi_s(j,\bar{r}_t)}{\Delta \bar{r}^2_*}=0,
\end{eqnarray}
and the iterative formula is
\begin{equation}\label{eq:iterative-upt}
\begin{split}
\Phi_s(j+1,k)=&-\Phi_s(j-1,k)+\left[2-2\frac{\Delta \bar{t}^2}{\Delta \bar{r}_*^2}-\Delta \bar{t}^2\bar{V}_s(k)\right]\Phi_s(j,k)\\
&-\frac{\Delta \bar{t}^2}{\Delta \bar{r}^2_*}C_s(\bar{a})(\bar{r}_{*k+1}-\bar{r}_t)\Phi_s(j,\bar{r}_t)
+\frac{\Delta \bar{t}^2}{\Delta \bar{r}_*^2}\left[\Phi_s(j,k+1)+\Phi_s(j,k-1)\right],
\end{split}
\end{equation}
with
\begin{equation}
    \Phi_s(j,\bar{r}_t)=\frac{(\bar{r}_t-\bar{r}_{*k})\Phi_s(j,k+1)-(\bar{r}_t-\bar{r}_{*k+1})\Phi_s(j,k)}{\Delta \bar{r}_*-C_s(\bar{a})(\bar{r}_t-\bar{r}_{*k})(\bar{r}_t-\bar{r}_{*k+1})}.
\end{equation}

In terms of the above iterative formulas, we can compute numerically perturbation waveforms  and QNMs, i.e., wave functions and QNM frequencies.
Here the corresponding boundary conditions are ingoing waves in $\bar{r}_*\to-\infty$ and outgoing waves in $\bar{r}_*\to\infty$.
In subsequent calculations, we take the initial wave packet as a Gaussian wave packet and the initial condition as
\begin{equation}\label{eq:initial}
\Phi_s(\bar{t}=0,\bar{r}_*)=e^{-\frac{(\bar{r}_*-A)^2}{2B^2}},\qquad \Phi_s(\bar{t}<0, \bar{r}_*)=0,
\end{equation}
where $A$ is the center of Gaussian packets, and $B$ the width.
In the present work, $B=1$ is set and the value of $A$ is chosen accordingly.
Moreover, considering the von Neumann stability condition~\cite{Zhu:2014sya},
\begin{equation}\label{eq:vonNeumann}
\frac{\Delta \bar{t}}{\Delta \bar{r}_*}<1,
\end{equation}
we take $\Delta \bar{t}/\Delta \bar{r}_*=0.25$.
According to the relationship~\cite{Berti:2009kk,Andersson:1995zk,Konoplya:2011qq} between wave functions $\Phi_s(\bar{t})$ and frequencies $\omega_j$ of QNMs,  
\begin{equation}\label{eq:Prony}
\Phi_s(\bar{t})\simeq \sum^p_{j=1}C_je^{-i 2M\omega_j\bar{t}},
\end{equation}
where $C_j$'s are excitation coefficients of $p$ main QNMs,
we extract $p$ main modes from  $\Phi_s(\bar{t})$ by using the Prony method~\cite{Konoplya:2011qq,Berti:2007dg} in the following three sections for the scalar, vector, and axial tensor field perturbations, respectively.
Among these $p$ QNMs, the dominant one that has the largest $|C_j|$ provides the most contributions to waveforms.

\section{Scalar field perturbation}\label{sec:scalar}
In this section we present the result that the waveform of scalar field perturbations contains two types: one type includes both the echo and the damping oscillation around Schwarzschild traversable wormholes, while the other type includes only the damping oscillation around Schwarzschild black holes. 

\subsection{Echo waveform}\label{sec:scalar_Echo}
When $\bar{a}>1$ and $\bar{a}-1\ll1$, the perturbation waveform around Schwarzschild traversable wormholes carries an echo waveform, i.e., the waveform  exhibits periodic pulse-shaped enhancements.
In the effective potential of scalar field perturbations at the throat, the coefficient of the $\delta$-function term is $C_{s=0}(\bar{a})=2(\bar{a}-1)/\bar{a}^2$, see Eq.~\eqref{csa}.
When the echo waveform occurs, we notice that $C_{s=0}(\bar{a})\ll 1$ due to $\bar{a}-1\ll1$.
In specific numerical simulations of waveforms, we find that there is no significant difference between the results obtained by taking $C_{s=0}(\bar{a})=2(\bar{a}-1)/\bar{a}^2$ and those obtained by taking $C_{s=0}(\bar{a})=0$.
Therefore, for scalar field perturbations, the influence of the effective potential at the throat on the waveform can be ignored.
We thus conclude that the echo waveform of scalar field perturbations is almost unaffected by the matter at the throat of the wormhole.

\begin{figure}[htbp]
	\centering
	\includegraphics[width=0.5\linewidth]{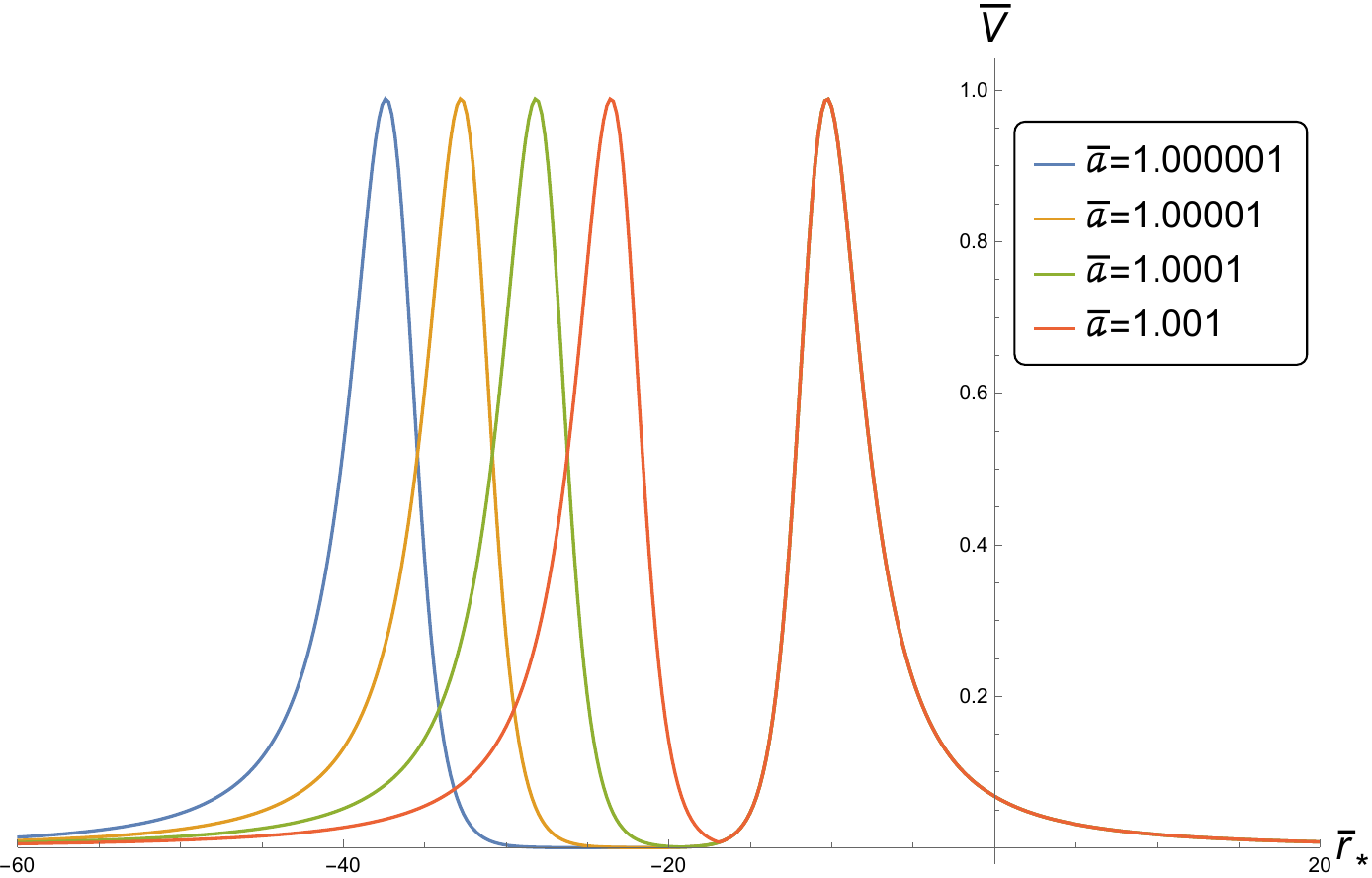}
	\caption{The effective potentials that result in the presence of echo waveforms under the scalar field perturbation with $l=2$ are shown. The blue, orange, green, and red curves correspond to $\bar{a}=1.000001, 1.00001, 1.0001$, and $1.001$, respectively. The right barriers of these four curves overlap.}
	\label{fig:potential-scalar}
\end{figure}

The generation of echoes is closely related to the shape of effective potentials under scalar field perturbations.
The effective potentials for a varying $\bar{a}$, as shown in Eq.~\eqref{eq:scalar-ot-dl}, result in the presence of echo waveforms and are depicted in Fig.~\ref{fig:potential-scalar}. 
Each of the four effective potentials exhibits two barriers: the left barrier and the right barrier. 
The four left barriers are separated, while the four right barriers overlap under different values of the parameter $\bar{a}$.
The reason for this phenomenon is that the Schwarzschild traversable wormhole is formed when the Schwarzschild spacetime, used as a template, is cut off at the throat ($\bar{\chi}=\bar{a}$) and then the ``mirror world" of the remaining spacetime is glued.
In other words, the effective potential of Schwarzschild traversable wormholes consists of two parts: the first part is the effective potential of the Schwarzschild spacetime cut off at the throat and the second part is the mirror of the first part, i.e., the first part is mapped to the observer’s universe.
Since the effective potential of the Schwarzschild spacetime is independent of the parameter $\bar{a}$, its potential barrier does not change with $\bar{a}$ and exhibits overlapping potential barriers under different values of $\bar{a}$, see the rightmost barrier in Fig.~\ref{fig:potential-scalar}.
The symmetry axis of effective potentials corresponds to the position of the throat of Schwarzschild traversable wormholes, namely, $\bar{r}_*=\bar{a}+\ln(\bar{a}-1)$.
Fig.~\ref{fig:potential-scalar} illustrates that echoes can only be generated in a double-barrier potential with a sufficiently large interval between the two barrier peaks, and such an interval acts as a resonant cavity.

\begin{figure}[htbp]
	\centering
	\subfigure[$\bar{a}=1.000001$]{
		\begin{minipage}[t]{0.4\linewidth}
			\centering
			\includegraphics[width=1\linewidth]{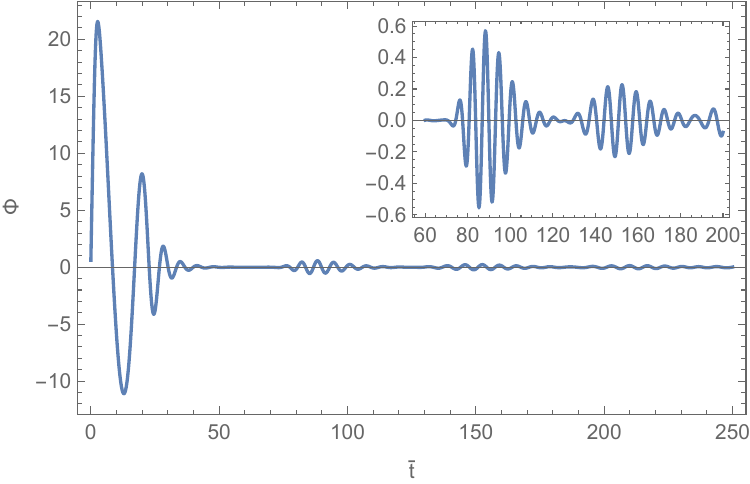}
		\end{minipage}
	}
	\subfigure[$\bar{a}=1.000001$]{
		\begin{minipage}[t]{0.4\linewidth}
			\centering
			\includegraphics[width=1\linewidth]{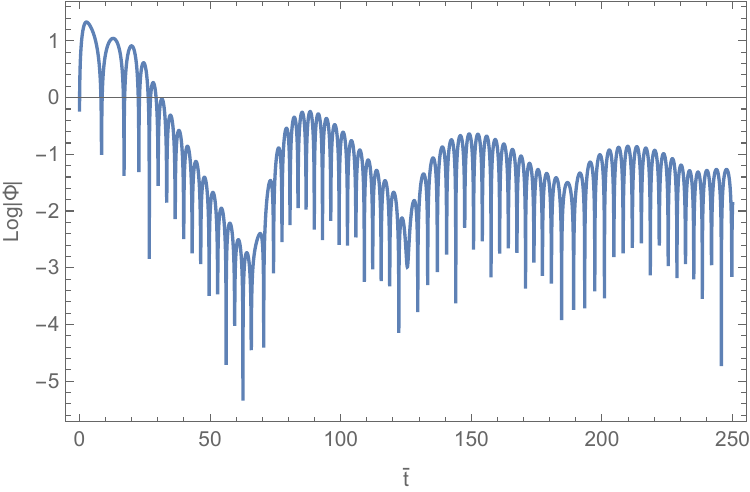}
		\end{minipage}
	}
	\vspace{-3mm}
	\subfigure[$\bar{a}=1.00001$]{
		\begin{minipage}[t]{0.4\linewidth}
			\centering
			\includegraphics[width=1\linewidth]{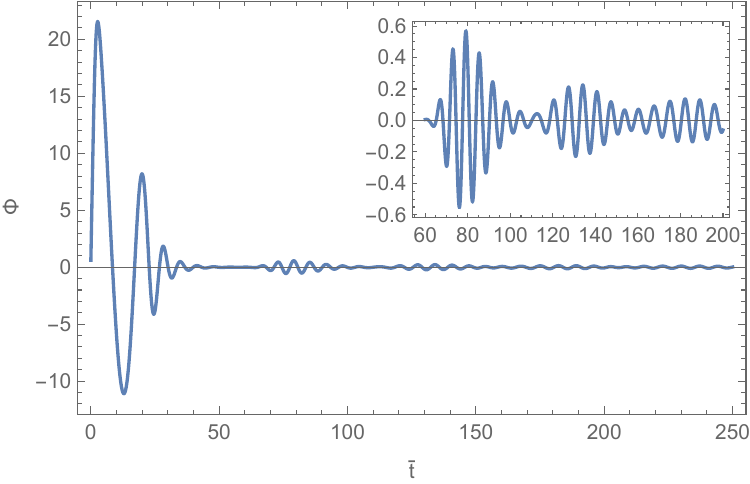}
		\end{minipage}
	}
	\subfigure[$\bar{a}=1.00001$]{
		\begin{minipage}[t]{0.4\linewidth}
			\centering
			\includegraphics[width=1\linewidth]{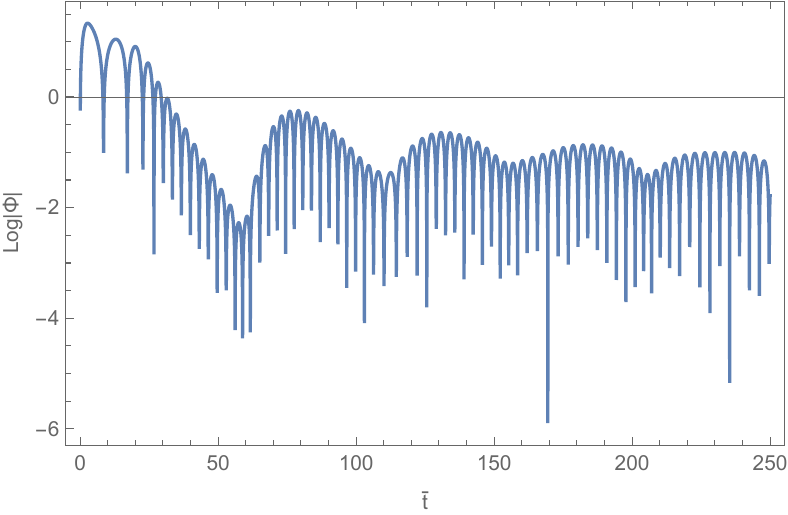}
		\end{minipage}
	}
	\vspace{-3mm}
	\subfigure[$\bar{a}=1.0001$]{
		\begin{minipage}[t]{0.4\linewidth}
			\centering
			\includegraphics[width=1\linewidth]{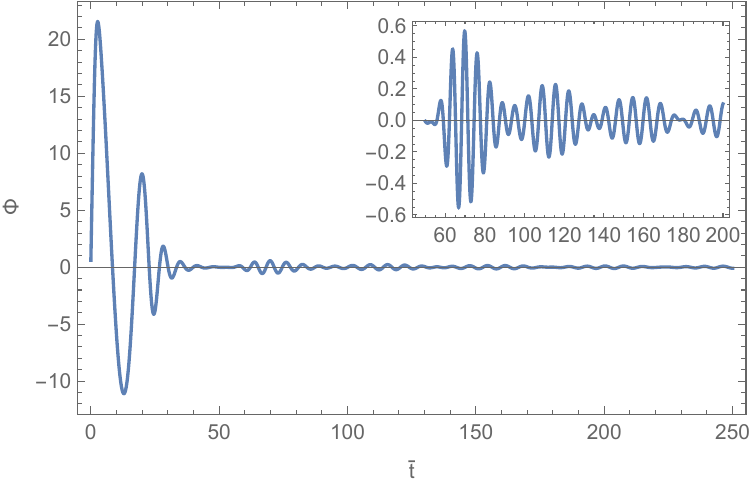}
		\end{minipage}
	}
	\subfigure[$\bar{a}=1.0001$]{
		\begin{minipage}[t]{0.4\linewidth}
			\centering
			\includegraphics[width=1\linewidth]{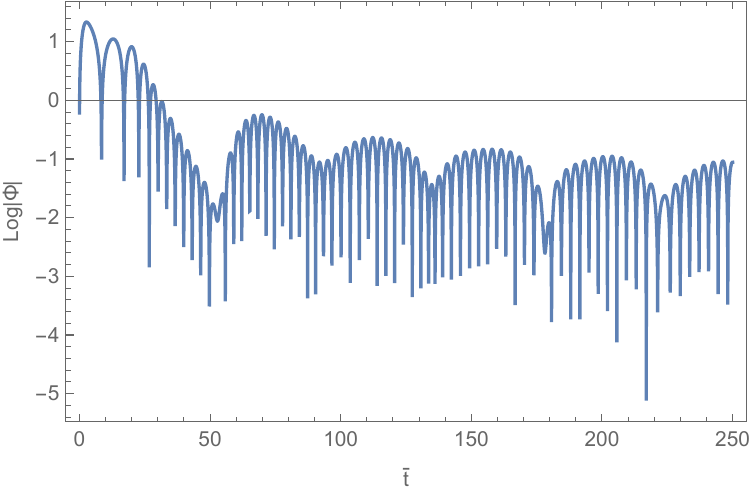}
		\end{minipage}
	}
	\vspace{-3mm}
	\subfigure[$\bar{a}=1.001$]{
		\begin{minipage}[t]{0.4\linewidth}
			\centering
			\includegraphics[width=1\linewidth]{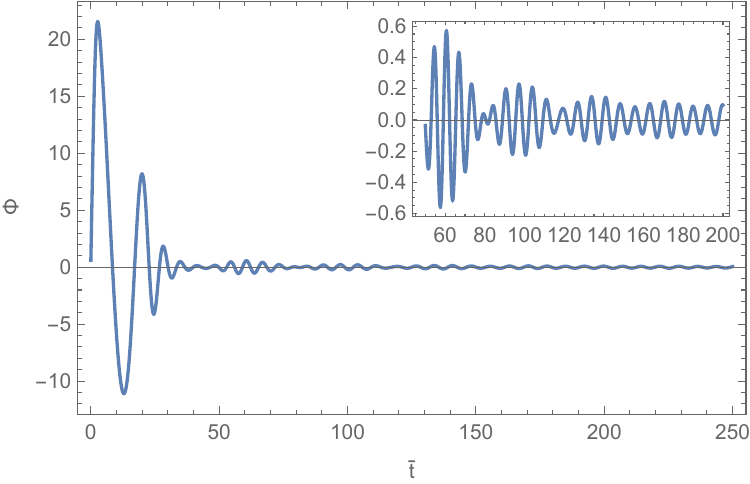}
		\end{minipage}
	}
	\subfigure[$\bar{a}=1.001$]{
		\begin{minipage}[t]{0.4\linewidth}
			\centering
			\includegraphics[width=1\linewidth]{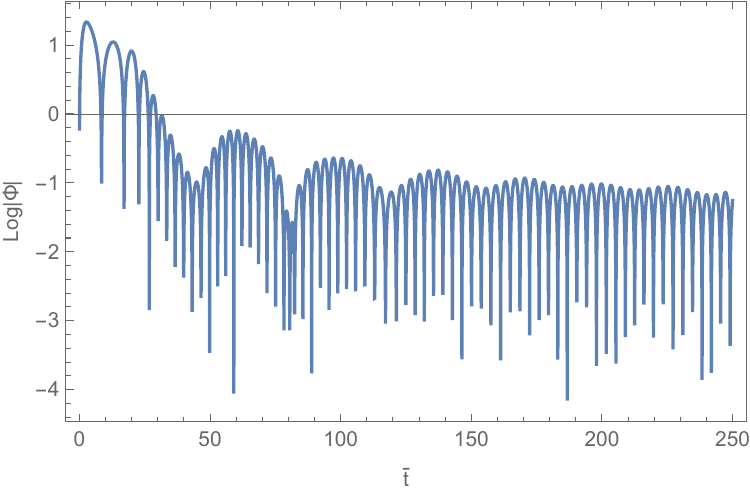}
		\end{minipage}
	}
	\caption{The echo waveforms around Schwarzschild traversable wormholes for different values of the parameter $\bar{a}$ under the scalar field perturbation with $l=2$. The left diagrams show how the wave function $\Phi$ varies with time $\bar{t}$, and the right ones show how the function $\log|\Phi|$ varies with time $\bar{t}$.}
	\label{fig:echo}
\end{figure}

By using the finite difference method and the Prony method introduced in Sec.~\ref{sec:FDM}, we plot the echo waveforms around Schwarzschild traversable wormholes in Fig. \ref{fig:echo} for different values of the parameter $\bar{a}$, and we also list the times when the first three echoes occur and the peaks of these echoes in Table \ref{Tab:1} under different values of the parameter $\bar{a}$.
We conclude that the echo waveforms around Schwarzschild traversable wormholes have the following characteristics:
\begin{itemize}
   \item For a fixed parameter $\bar{a}$, the three echoes appear intermittently and their peaks gradually decrease from the first to the third due to wave scattering by potential barriers during propagation.
   \item Under different values of the parameter $\bar{a}$, the peaks of echoes are almost equal because the peaks of the potential barriers are exactly the same.
   \item  When the parameter $\bar{a}$ increases, the time at which each echo occurs becomes earlier and the time interval between adjacent echoes becomes shorter because the distance between the two potential barriers decreases.
\end{itemize}

\begin{table}[!ht]
	\centering
	\caption{This table shows the times at which the first three echoes occur and the peaks of these echoes under different values of the parameter $\bar{a}$ for scalar field perturbations.}
	\begin{tabular}{cccccccc}
		\hline 
		
		$\bar{a}$     & Echo 1 (T) & Echo 1 (P) & Echo 2 (T) & Echo 2 (P)& Echo 3 (T) &Echo 3 (P)\\ \hline
		
		$1.000001$         & 88.30      & 0.567498   & 149.20     & 0.228950  & 209.90     &0.138538  \\ \hline
		$1.00001$          & 79.10      & 0.567204   & 130.75     & 0.227911  & 182.20     &0.136524  \\ \hline
		$1.0001$           & 69.85      & 0.566514   & 112.25     & 0.231643  & 158.00     &0.146416  \\ \hline
		$1.001$            & 60.65      & 0.568974   &  97.25     & 0.228812  & 137.35     &0.152663  \\ \hline
	\end{tabular}
	\label{Tab:1}
\end{table}

In summary, the scalar field perturbation generates an echo waveform in the wormhole spacetime when the conditions $\bar{a}>1$ and $\bar{a}-1\ll1$ are satisfied.
From the discussion in Sec.~\ref{sec:Schwarzschild sol}, we can see that the appearance of an exotic compact object as a black hole or a wormhole is closely related to its mass $M$.
Specifically, when $M > a/2$, the exotic compact object behaves as a black hole; when $M = a/2$, it behaves as a one-way wormhole; and when $M < a/2$, it behaves as a traversable wormhole.
Therefore, when the mass decreases from more than $a/2$ to equal to or less than $a/2$, the spacetime structure of a black hole changes into that of a wormhole, which may be achieved through Hawking radiation. 
When the mass increases from equal to or less than $a/2$ to more than $a/2$, the spacetime structure of a wormhole changes into that of a black hole, which may be achieved through the wormhole throat absorbing matter and increasing its mass.
According to Eqs.~\eqref{r*WH-dl} and \eqref{eq:scalar-ot-dl}, the effective potential barrier at the observer’s universe does not change with $\bar{a}$.
When $\bar{a}$ increases, the position of the wormhole throat, i.e. $\bar{r}_*|_{\bar{\chi}=\bar{a}}=\bar{a}+\ln(\bar{a}-1)$, increases, resulting in a decrease in the distance between the throat and the effective potential barrier at the observer’s universe.
Since the two potential barriers are symmetric with respect to the wormhole throat, the increase in $\bar{a}$ will reduce the interval between the effective potential barrier at the observer’s universe and its mirror barrier at the opposite universe.
Therefore, echoes occur more frequently for a larger $\bar{a}$. 
However, as is known, the scalar field perturbation only produces a damping oscillation waveform without echoes in the case of Schwarzschild black holes.

\subsection{Damping oscillation}\label{sec:S-Damping}
\begin{figure}[htbp]
	\centering
	\includegraphics[width=0.8\linewidth]{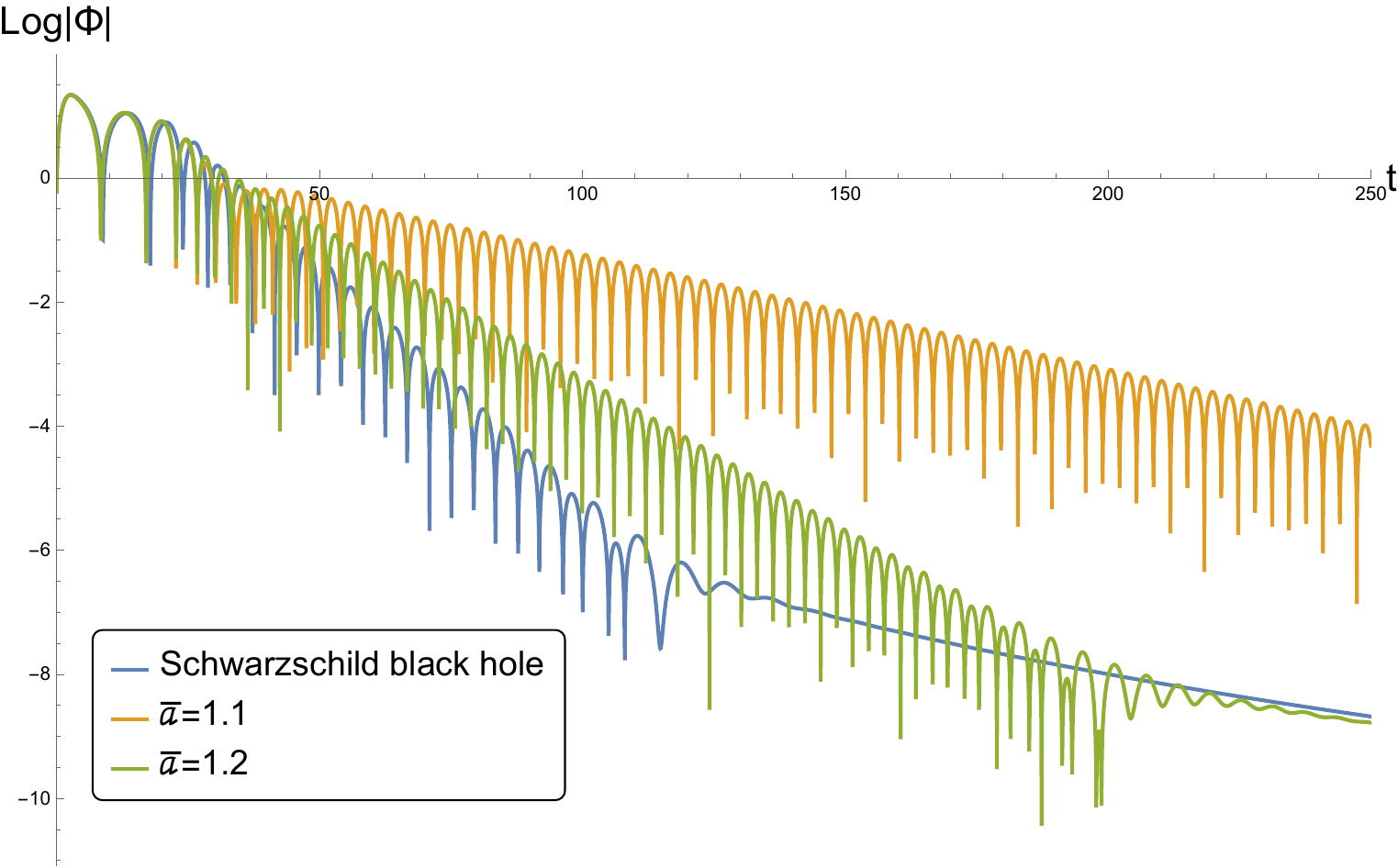}
	\caption{This figure shows the damping oscillation waveforms under the scalar field perturbation with  $l=2$ for different values of the parameter $\bar{a}$, where the blue curve corresponds to Schwarzschild black holes as a comparison, while the orange and green curves correspond Schwarzschild traversable wormholes in the cases of $\bar{a}=1.1, 1.2$, respectively.}
	\label{fig:QNM}
\end{figure}

As $\bar{a}$ becomes large, the perturbation waveform gradually changes from echoes to damping oscillations.
Here we start our analysis from $\bar{a}\geq1.1$, where only the damping oscillation waveform exists in the spacetime.
The characteristic frequency of damping oscillations is described by the QNM, which consists of a real part and an imaginary part.
Here we again use the finite difference method and the Prony method introduced in Sec.~\ref{sec:FDM} to calculate the waveform and QNMs in the parameter range of $\bar{a}\geq1.1$.
The dominant QNMs are given in Table~\ref{Tab:QNM_Scalar-delta}, where the data are dimensionless because the frequency $\omega$ has been multiplied by the factor $2M$, and the relation between the logarithm wave function $\log|\Phi|$ and the dimensionless time $\bar{t}$ is depicted in Fig.~\ref{fig:QNM}.

\begin{table}[!ht]
	\centering
	\caption{This table shows the types of waveforms and QNMs under the scalar field perturbation with $l=2$. Here, the case of $\bar{a}\leq 1$  corresponds to Schwarzschild black holes, the case of $1.000001\leq\bar{a}\leq 1.001 $ corresponds to Schwarzschild traversable wormholes with obvious echoes, and the case of $1.1\leq\bar{a}\leq 10.0 $ corresponds to Schwarzschild traversable wormholes with only damping oscillations.}
	\begin{tabular}{cccc}\hline
		
		$\bar{a}$          & $2M\omega$ (waveforms/QNMs) & $\bar{a}$          & $2M\omega$ (waveforms/QNMs)\\ \hline
		
		$\leq 1$ (black holes)          & 0.967508 - 0.193780 i   &$1.7$              & 0.923455 - 0.261942 i\\ \hline
		$1.000001$         & Echo                                 &$1.8$              & 0.889434 - 0.269808 i\\ \hline
		$1.00001$          & Echo                                 &$1.9$              & 0.856480 - 0.274453 i\\ \hline
		$1.0001$           & Echo                                 &$2.0$              & 0.825027 - 0.276598 i \\ \hline
		$1.001$            & Echo                                 &$3.0$              & 0.591876 - 0.249052 i\\ \hline
		$1.01$             & 0.916444 - 0.012698 i (Weak echo)    &$4.0$              & 0.456868 - 0.210061 i\\ \hline
		$1.1$              & 0.974884 - 0.042691 i                &$5.0$              & 0.371175 - 0.179118 i\\ \hline
		$1.2$              & 1.038524 - 0.112627 i                &$6.0$              & 0.312386 - 0.155342 i\\ \hline
		$1.3$              & 1.040700 - 0.166503 i                &$7.0$              & 0.270053 - 0.136100 i\\ \hline
		$1.4$              & 1.020510 - 0.204789 i                &$8.0$              & 0.237024 - 0.122119 i\\ \hline
		$1.5$              & 0.990816 - 0.231557 i                &$9.0$              & 0.205958 - 0.095188 i\\ \hline
		$1.6$              & 0.957772 - 0.249718 i                &$10.0$             & 0.177493 - 0.065322 i\\ \hline
	\end{tabular}
	\label{Tab:QNM_Scalar-delta}
\end{table}

In Table~\ref{Tab:QNM_Scalar}, we present the QNMs of waveforms after ignoring the influence of the matter at the wormhole throat, i.e., taking $C_{s=0}(\bar{a})=0$.
It can be seen that, under the same parameter $\bar{a}$, the absolute values of the real and imaginary parts of the QNMs in Table~\ref{Tab:QNM_Scalar-delta} are both greater than those in Table~\ref{Tab:QNM_Scalar}. 
This indicates that the matter at the wormhole throat intensifies the oscillation frequency of the scalar field perturbation in the spacetime and accelerates the decay rate of the perturbation.

\begin{table}[!ht]
	\centering
	\caption{This table shows the types of waveforms and QNMs under the scalar field perturbation with $l=2$ when the influence of matter at the wormhole throat is ignored, i.e. $C_{s=0}(\bar{a})=0$. Here, the case of $\bar{a}\leq 1$  corresponds to Schwarzschild black holes, the case of $1.000001\leq\bar{a}\leq 1.001 $ corresponds to Schwarzschild traversable wormholes with obvious echoes, and the case of $1.1\leq\bar{a}\leq 10.0 $ corresponds to Schwarzschild traversable wormholes with only damping oscillations.}
	\begin{tabular}{cccc}\hline
		
		$\bar{a}$          & $2M\omega$ (waveforms/QNMs) & $\bar{a}$          & $2M\omega$ (waveforms/QNMs)\\ \hline
		
		$\leq 1$ (black holes)          & 0.967508 - 0.193780 i   &$1.7$              & 0.881462 - 0.194273 i\\ \hline
		$1.000001$         & Echo                    &$1.8$              & 0.847602 - 0.198093 i\\ \hline
		$1.00001$          & Echo                    &$1.9$              & 0.814952 - 0.199712 i\\ \hline
		$1.0001$           & Echo                    &$2.0$              & 0.783926 - 0.199760 i \\ \hline
		$1.001$            & Echo                    &$3.0$              & 0.556853 - 0.171306 i\\ \hline
		$1.01$             & 0.916457 - 0.012704 i (Weak echo)& $4.0$              & 0.428506 - 0.141430 i\\ \hline
		$1.1$              & 0.952315 - 0.035264 i &$5.0$              & 0.347028 - 0.118807 i\\ \hline
		$1.2$              & 1.007350 - 0.090619 i &$6.0$              & 0.291499 - 0.103299 i\\ \hline
		$1.3$              & 1.004790 - 0.131173 i &$7.0$              & 0.250827 - 0.090494 i\\ \hline
		$1.4$              & 0.981648 - 0.158289 i &$8.0$              & 0.218958 - 0.079519 i\\ \hline
		$1.5$          & 0.950439 - 0.176095 i &$9.0$              & 0.193980 - 0.072122 i\\ \hline
		$1.6$              & 0.916186 - 0.187382 i &$10.0$             & 0.184465 - 0.069729 i\\ \hline
	\end{tabular}
	\label{Tab:QNM_Scalar}
\end{table}

For a stationary observer, since the contribution of dominant modes is much larger than that of the other modes, the observed waveform approximately equals the dominant quasi-normal mode,
\begin{equation}\label{QNM_Re}
\Phi\sim C \cdot e^{-i\omega t}=C \cdot e^{-i\cdot2M\omega \bar{t}},
\end{equation}
where $C$ is the quasi-normal excitation coefficient corresponding to the dominant mode $\omega$.
Next, we give the difference of $\bar{t}$ between two adjacent peaks in the damping oscillation region on the $\log|\Phi|-\bar{t}$ plot,
\begin{equation}\label{Delta_ht}
\Delta \bar{t}=\frac{\pi}{2M\omega_{\rm R}},
\end{equation}
and the ordinate difference between two adjacent peaks,
\begin{equation}\label{eq:Delta-Log}
\Delta \log|\Phi|=\frac{\pi\omega_{\rm I}}{\omega_{\rm R}}.
\end{equation}
As a result, we obtain the slope of the line connecting two adjacent peaks,
\begin{equation}\label{slope}
\bar{k}=\frac{\Delta \log|\Phi|}{\Delta \bar{t}}=2M\omega_{\rm I}.
\end{equation}

\begin{figure}[htbp]
	\centering
	\subfigure[The real part of QNMs]{\label{fig:QNM-1-a}
		\begin{minipage}[t]{0.45\linewidth}
			\centering
			\includegraphics[width=1\linewidth]{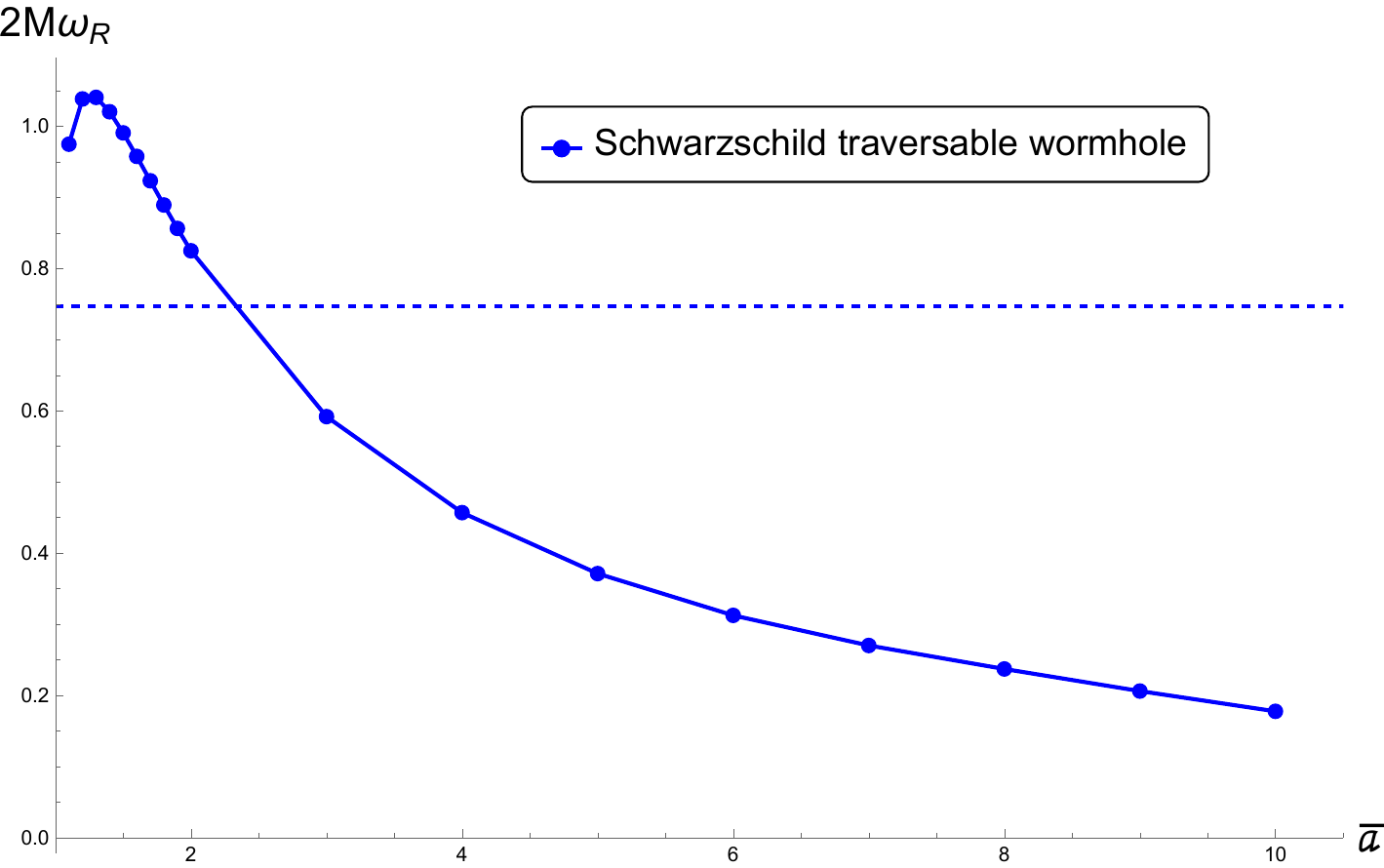}
		\end{minipage}
	}
	\subfigure[The imaginary part of QNMs]{\label{fig:QNM-1-b}
		\begin{minipage}[t]{0.45\linewidth}
			\centering
			\includegraphics[width=1\linewidth]{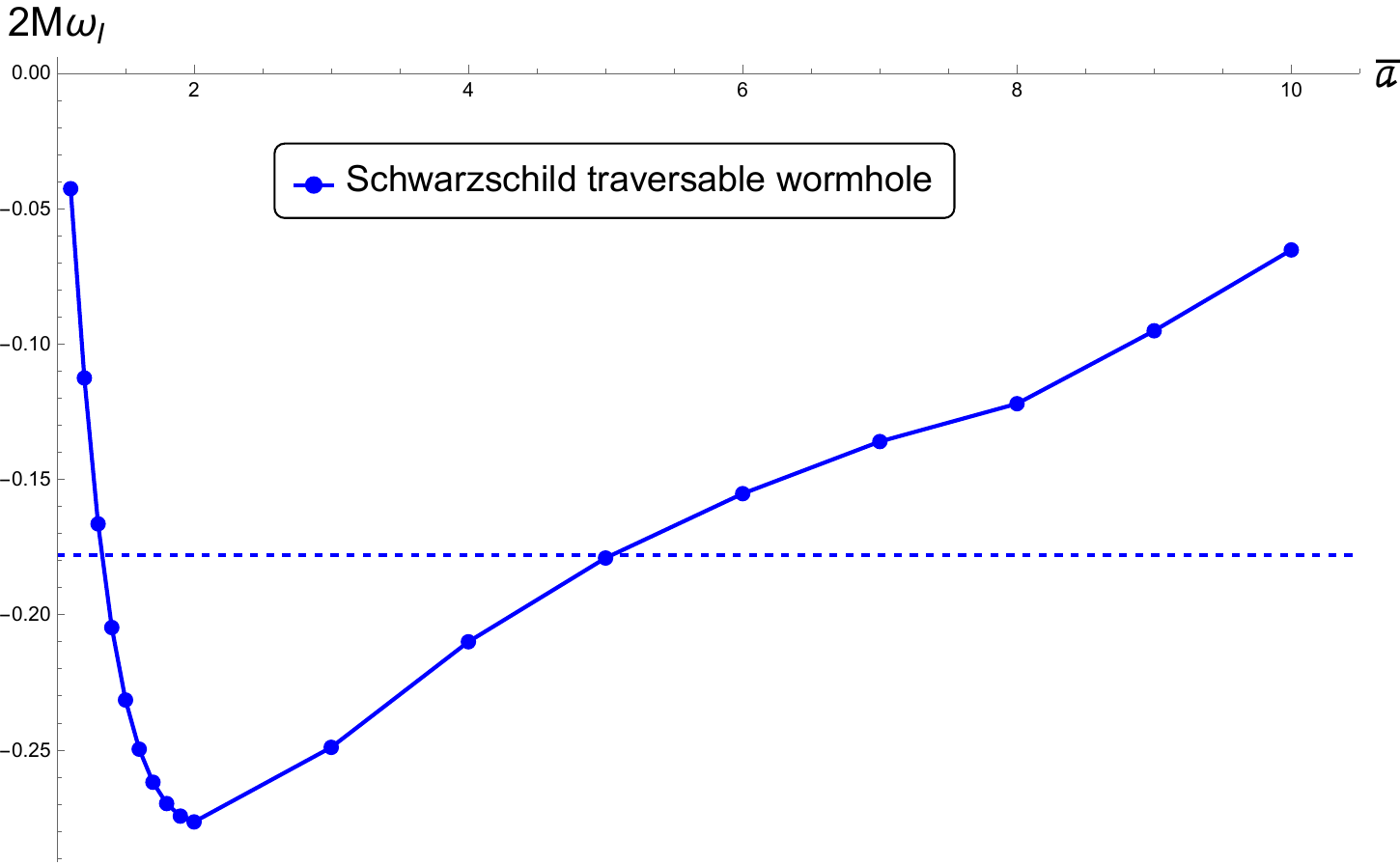}
		\end{minipage}
	}
	\caption{The two plots show how the real and imaginary parts of QNMs for Schwarzschild traversable wormholes vary with the parameter $\bar{a}$ under the scalar field perturbation with $l=2$.  
    The solid curves are the corresponding variation curves, where the dots represent relevant data given by Table~\ref{Tab:QNM_Scalar-delta}.
    The dotted lines are constant reference lines parallel to the $\bar{a}$ axis, and their fixed vertical coordinate values correspond to the real and imaginary parts of QNMs for the Schwarzschild black hole.
    These fixed values are used to compare with the vertical coordinate values of the solid curves.}
	\label{fig:QNM-1}
\end{figure}

By employing the Prony method to calculate the QNMs for various values of $\bar{a}$ in Schwarzschild traversable wormholes, we present the plots of $2M\omega_{\text{R}}$ versus $\bar{a}$ and $2M\omega_{\text{I}}$ versus $\bar{a}$ in Fig.~\ref{fig:QNM-1-a} and Fig.~\ref{fig:QNM-1-b}, respectively. 
The dots in these figures correspond to the data listed in Table~\ref{Tab:QNM_Scalar-delta}.
As shown in Table~\ref{Tab:QNM_Scalar-delta},  $2M\omega_{\rm R}$ and $2M\omega_{\rm I}$ are constants for Schwarzschild black holes, but vary with $\bar{a}$ for Schwarzschild traversable wormholes.
Therefore, $\Delta \bar{t}, \bar{k}$ and $\Delta\log|\Phi|$ remain unchanged for Schwarzschild black holes, but vary with $\bar{a}$ for Schwarzschild traversable wormholes in the $\log|\Phi|-\bar{t}$ plot.
By comparing the differences in $\Delta \bar{t}$, $\bar{k}$, and $\Delta\log|\Phi|$  between these two types of spacetime, we are able to distinguish Schwarzschild black holes from Schwarzschild traversable wormholes.
To this end, we need the physical time $t$ and the other quantities with dimension, such as $k$.
According to the relationship in Eq.~\eqref{eq:dimensionless} between the dimensionless time and the physical time, the physical time interval and the corresponding slope between two adjacent peaks take the forms,
\begin{equation}\label{eq:dt-hdt}
\Delta t=2M\Delta\bar{t},\qquad  k=\frac{\bar{k}}{2M}.
\end{equation}
We can see that the two quantities vary with $M$ for Schwarzschild black holes.
Since we have no data on the mass parameter $M$, we cannot distinguish Schwarzschild black holes from Schwarzschild traversable wormholes by using $\Delta t$ and $k$.
However, the quantity $\Delta\log|\Phi|$ remains unchanged with respect to $M$ for Schwarzschild black holes, but varies with $\bar{a}$ for Schwarzschild traversable wormholes.
Therefore, we can use the value of $\Delta\log|\Phi|$ to distinguish these two spacetimes.
According to the QNM of Schwarzschild black holes, $2M\omega=0.967508 - 0.193780 i$ (see the second row in Table~\ref{Tab:QNM_Scalar-delta}), and Eq.~\eqref{eq:Delta-Log}, we obtain
\begin{equation}\label{eq:Delta_Phi-SCH}
\Delta\log|\Phi|=\frac{-0.193780}{0.967508}\pi\approx -0.629223.
\end{equation}
If the difference of $\log|\Phi|$ between adjacent peaks in damping oscillations deviates significantly from Eq.~\eqref{eq:Delta_Phi-SCH}, we can infer that the spacetime is not a Schwarzschild black hole, but rather a Schwarzschild traversable wormhole.

\subsection{Scenario for fixing parameters}\label{sec:method}
\begin{figure}[htbp]
\centering

\includegraphics[width=0.5\linewidth]{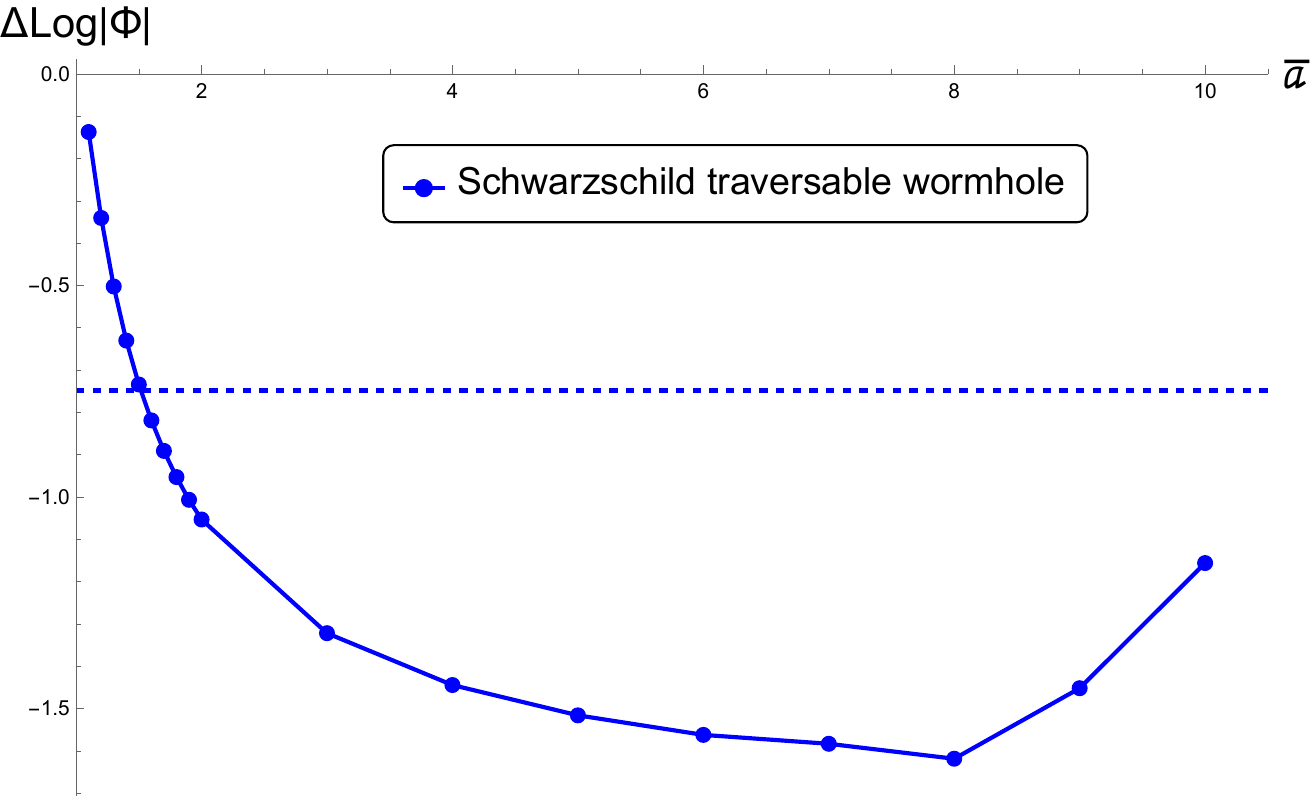}

\caption{This figure shows how the difference of $\log|\Phi|$ between adjacent peaks,  $\Delta\log|\Phi|$, for Schwarzschild traversable wormholes varies with the parameter $\bar{a}$ under the scalar field perturbation with $l=2$. 
The solid curve is the corresponding variation curve, where the dots represent relevant data given by Table~\ref{Tab:QNM_Scalar-delta}.
The dotted line is a constant reference line parallel to the $\bar{a}$ axis, and its fixed vertical coordinate value corresponds to  $\Delta\log|\Phi|$ of the Schwarzschild black hole.
This fixed value is used to compare with the vertical coordinate values of the solid curve.}
\label{fig:QNM-2}
\end{figure}

In the experimental observations~\cite{LIGOScientific:2016aoc,LIGOScientific:2017vwq}, the waveform data are depicted in a $\Phi-t$ plot, and then the corresponding $\log|\Phi|-t$ plot can be obtained through data processing.
In the $\log|\Phi|-t$ plot, we can extract two parameters: the time interval between adjacent peaks, $\Delta t$, and the difference of $\log|\Phi|$ between adjacent peaks, $\Delta\log|\Phi|$.
With these two parameters, we can determine the parameters $M$ and $a$ for Schwarzschild traversable wormholes by following the three steps:
\begin{itemize}
\item By fixing the value of $\Delta\log|\Phi|$, we give the value of $\bar{a}$ in Fig.~\ref{fig:QNM-2}, which is a theoretical plot drawn up in accordance with Eq.~\eqref{eq:Delta-Log} and Fig.~\eqref{fig:QNM-1};
\item By using the value of $\bar{a}$, we obtain $2M\omega_{\rm R}$ as shown in Fig.~\ref{fig:QNM-1};
\item By using Eqs.~\eqref{Delta_ht} and \eqref{eq:dt-hdt}, together with the value of  $\Delta t$, we obtain the mass parameter,
\begin{equation}\label{eq:method_M}
M=\frac{\Delta t}{2\Delta\bar{t}}=\frac{(2M\omega_{\rm R})\Delta t}{2\pi},
\end{equation}
and then the parameter $a$ by combining $M$ and $\bar{a}$ as follows,
\begin{equation}\label{eq:method_a}
a=2M\bar{a}.
\end{equation}
\end{itemize}

\section{Electromagnetic field perturbation}\label{sec:EM}
\begin{figure}[htbp]
\centering
\includegraphics[width=0.8\linewidth]{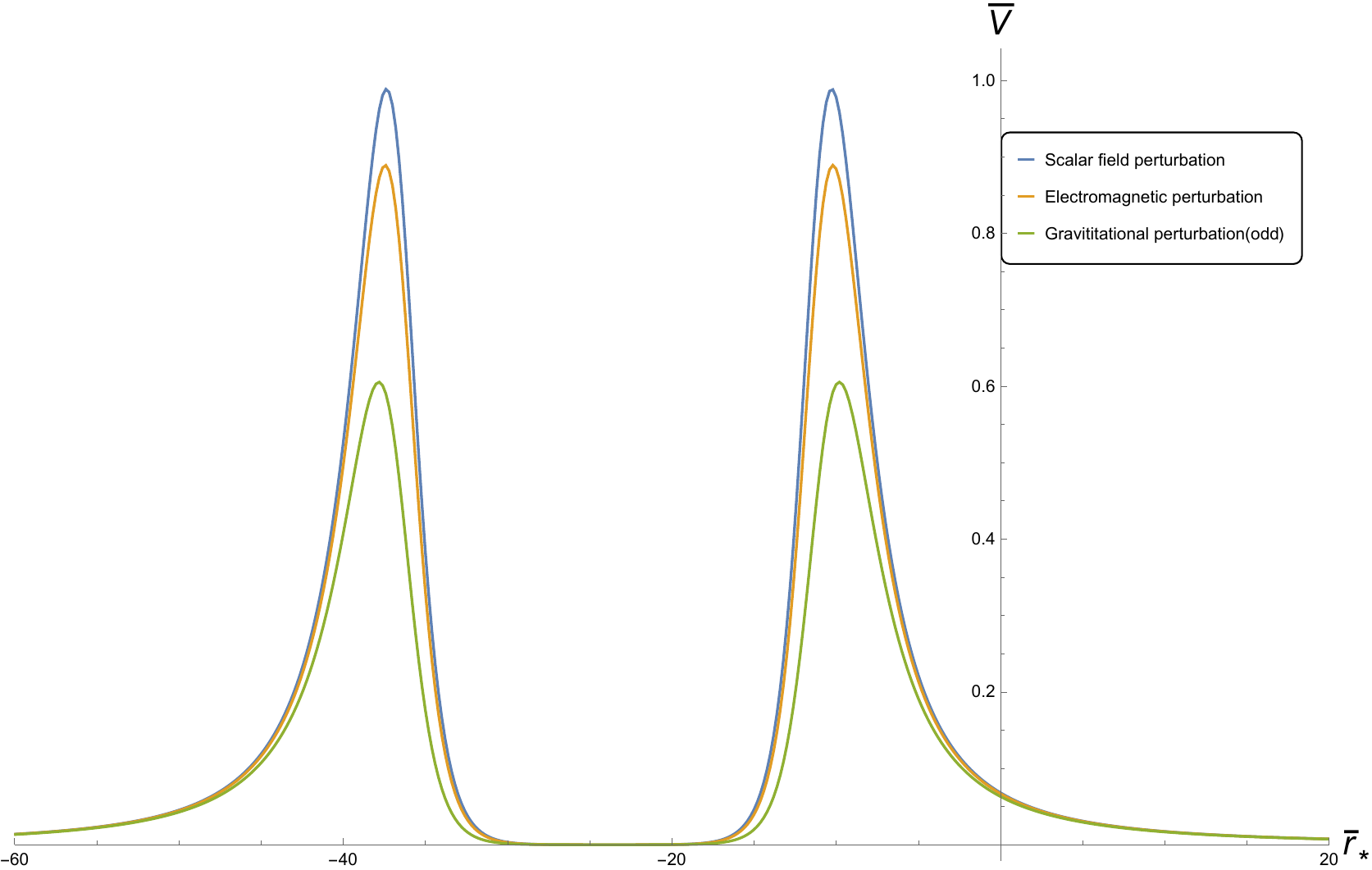}
\caption{The figure shows the effective potentials outside the throat of Schwarzschild traversable wormholes under various field perturbations, where $l=2$ and $\bar{a}=1.000001$ are set. The blue, orange and green curves correspond to the scalar field, electromagnetic field and axial gravitational field  perturbations, respectively.}
\label{fig:potential-all}
\end{figure}

The main difference between the electromagnetic field perturbation and the other two perturbations is that the former is completely unaffected by the matter at the throat of the wormhole.
However, the waveform types under the electromagnetic field perturbation are similar to those under the scalar field perturbation because the effective potential of the electromagnetic field perturbation has a similar shape to that of the scalar field perturbation outside the throat of Schwarzschild traversable wormholes, and the influence of the matter at the throat of the wormhole on the perturbation of the scalar field is negligible when the echo occurs.
For a more intuitive comparison, we present the effective potentials outside the throat of Schwarzschild traversable wormholes under various field perturbations in Fig.~\ref{fig:potential-all} by using Eqs.~\eqref{eq:EM-dl}-\eqref{eq:oddgravity-ot-dl},  where $l=2$ and $\bar{a}=1.000001$ are set.
However, the peak values of effective potentials are  different for different field perturbations, which will influence the QNMs and the peak values of echoes.

\subsection{Echo waveform}\label{sec:EM_Echo}
\begin{figure}[htbp]
\centering
\subfigure[$\bar{a}=1.000001$]{
    \begin{minipage}[t]{0.45\linewidth}
    \centering
    \includegraphics[width=1\linewidth]{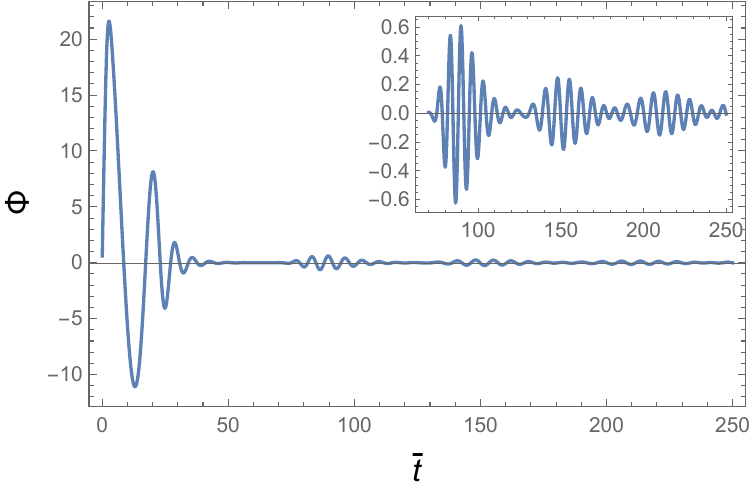}
    \end{minipage}
}
\subfigure[$\bar{a}=1.000001$]{
    \begin{minipage}[t]{0.45\linewidth}
    \centering
    \includegraphics[width=1\linewidth]{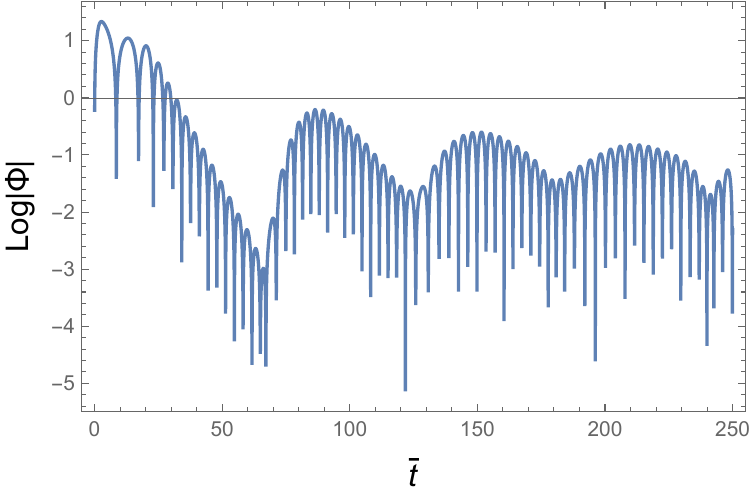}
    \end{minipage}
}

\caption{The echo waveform around Schwarzschild traversable wormholes for $\bar{a}=1.000001$ under the electromagnetic field perturbation with $l=2$. The left diagram shows how the wave function $\Phi$ varies with the time $\bar{t}$, and the right one shows how the function $\log|\Phi|$ varies with time $\bar{t}$.}
\label{fig:EM-echo}
\end{figure}

The waveform of electromagnetic field perturbations contains echoes when $\bar{a}>1$ and $\bar{a}-1\ll 1$.
Within this range, we use the finite difference method to calculate the waveform and extract the occurrence times and peak values of echoes.
In Fig.~\ref{fig:EM-echo}, we show the waveform under the electromagnetic field perturbation when $\bar{a}=1.000001$.
In Table~\ref{Tab:Echo-EM}, we list the times at which the first three echoes occur and the peak values of the echoes for different values of the parameter $\bar{a}$. 

\begin{table}[!ht]
    \centering
    \caption{This table shows the times at which the first three echoes occur and the peaks of these echoes under different values of the parameter $\bar{a}$ for electromagnetic field perturbations.}
  \begin{tabular}{cccccccc}
  \hline 
  
    $\bar{a}$     & Echo 1 (T) & Echo 1 (P) & Echo 2 (T) & Echo 2 (P)& Echo 3 (T) &Echo 3 (P)\\ \hline

    $1.000001$         & 86.4      & 0.620130   & 151.65    & 0.249708  & 209.75     &0.150211 \\ \hline
    $1.00001$          & 77.2     & 0.620059   & 133.2     & 0.249948  & 182.05     &0.154092  \\ \hline
    $1.0001$           & 67.95      & 0.620921  & 114.75    & 0.251282  & 161.70     &0.145785  \\ \hline
    $1.001$            & 58.75     & 0.623614   &  99.80     & 0.235031  &141.30    &0.166783  \\ \hline
  \end{tabular}
    \label{Tab:Echo-EM}
\end{table}

By comparing Table~\ref{Tab:1} with Table~\ref{Tab:Echo-EM}, we analyze the similarities and differences between the waveforms of scalar field perturbations and electromagnetic field perturbations.
It is similar to the case of scalar field perturbations that the variation of parameter $\bar{a}$ has little influence on the peak value of every order echo in the electromagnetic field perturbation.
Moreover, the peak value of each order echo for a fixed $\bar{a}$ under the electromagnetic field perturbation is higher than that of the same order echo under the scalar field perturbation.
The reason is that the peak value of effective potentials in the electromagnetic field perturbation is lower than that in the scalar field perturbation, which indicates that the former potential barrier has a weaker scattering effect on waves than the latter one, resulting in a higher echo peak for the former.

\subsection{Damping oscillation}\label{sec:EM-Damping}
When $\bar{a}$ increases, the waveform of electromagnetic field perturbations changes from echoes to damping oscillations, which is similar to that of scalar field perturbations.
In Table~\ref{Tab:QNM_EM}, we show the QNMs computed by the Prony method for different values of $\bar{a}$.
Comparing Table~\ref{Tab:QNM_EM} with Table~\ref{Tab:QNM_Scalar}, we observe that both the real parts and the absolute values of the imaginary parts under the electromagnetic field perturbation are smaller than those under the scalar field perturbation for a fixed $\bar{a}$. 
This behavior indicates that the oscillation period of electromagnetic field perturbations is longer than that of scalar field perturbations, and that the modes are more stable under electromagnetic field perturbations.
The reason is that the peak value of effective potentials is smaller in the electromagnetic field perturbation, which reduces the suppression effect of potential barriers on waveforms.

\begin{table}[!ht]
    \centering
    \caption{This table shows the waveform types and QNMs under the electromagnetic field perturbation with $l=2$. Here, the case of $\bar{a}\leq 1$ corresponds to Schwarzschild black holes, the case of $1.000001\leq\bar{a}\leq 1.001 $ corresponds to Schwarzschild traversable wormholes with obvious echoes, and the case of $1.1\leq\bar{a}\leq 10.0 $ corresponds to Schwarzschild traversable wormholes with only damping oscillations.}
  \begin{tabular}{cccc}\hline
  
    $\bar{a}$          & waveform/QNM ($2M\omega$) & $\bar{a}$          & waveform/QNM ($2M\omega$)\\ \hline
  
    $\leq 1$           & 0.915464 - 0.190100 i  &$1.7$              & 0.844303 - 0.185849 i\\ \hline
    $1.000001$   & Echo                            &$1.8$              & 0.813333 - 0.189897 i\\ \hline
    $1.00001$     & Echo                             &$1.9$              & 0.783507 - 0.191816 i\\ \hline
    $1.0001$       & Echo                             &$2.0$              & 0.754935 - 0.192228 i \\ \hline
    $1.001$          & Echo                             &$3.0$              & 0.542394 - 0.166944 i\\ \hline
    $1.01$            & 0.877937 - 0.014752 i (Weak echo)& $4.0$ & 0.419618 - 0.138658 i\\ \hline
    $1.1$              &0.900571 - 0.034332 i  &$5.0$              & 0.342186 - 0.117842 i\\ \hline
    $1.2$              & 0.953902 - 0.086300 i &$6.0$              & 0.287567 - 0.102703 i\\ \hline
    $1.3$              & 0.953680 - 0.124633 i &$7.0$              & 0.247981 - 0.089702 i\\ \hline
    $1.4$              & 0.933921 - 0.150537 i &$8.0$              & 0.213115 - 0.075908 i\\ \hline
    $1.5$              & 0.906326 - 0.167754 i &$9.0$              & 0.191102 - 0.068795 i\\ \hline
    $1.6$              & 0.875648 - 0.178906 i &$10.0$            & 0.172544 - 0.057280 i\\ \hline
  \end{tabular}
    \label{Tab:QNM_EM}
\end{table}

\begin{figure}[htbp]
\centering
\subfigure[The real part of QNMs]{\label{fig:QNMEM-1-a}
    \begin{minipage}[t]{0.45\linewidth}
    \centering
    \includegraphics[width=1\linewidth]{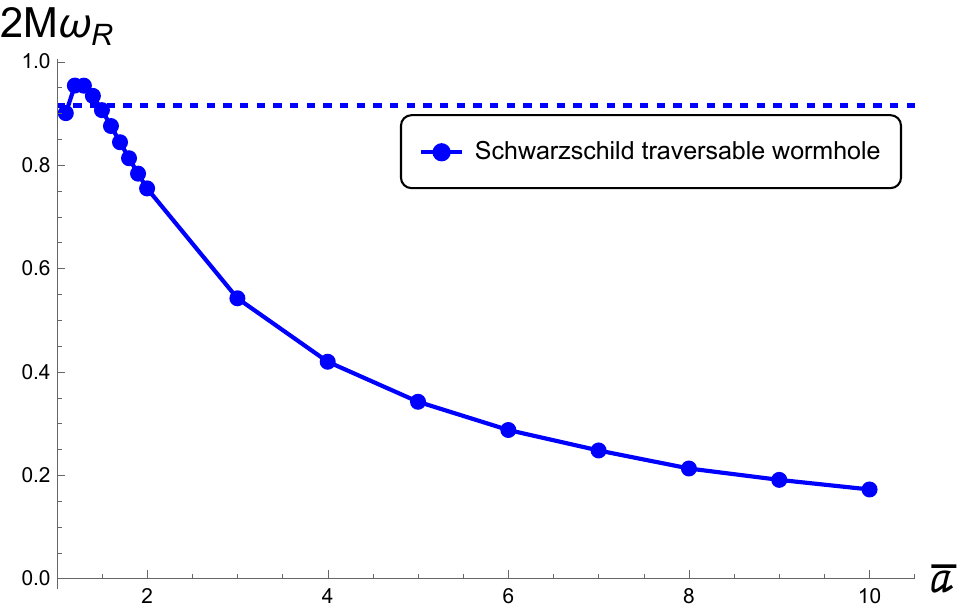}
    \end{minipage}
}
\subfigure[The imaginary part of QNMs]{\label{fig:QNMEM-1-b}
    \begin{minipage}[t]{0.45\linewidth}
    \centering
    \includegraphics[width=1\linewidth]{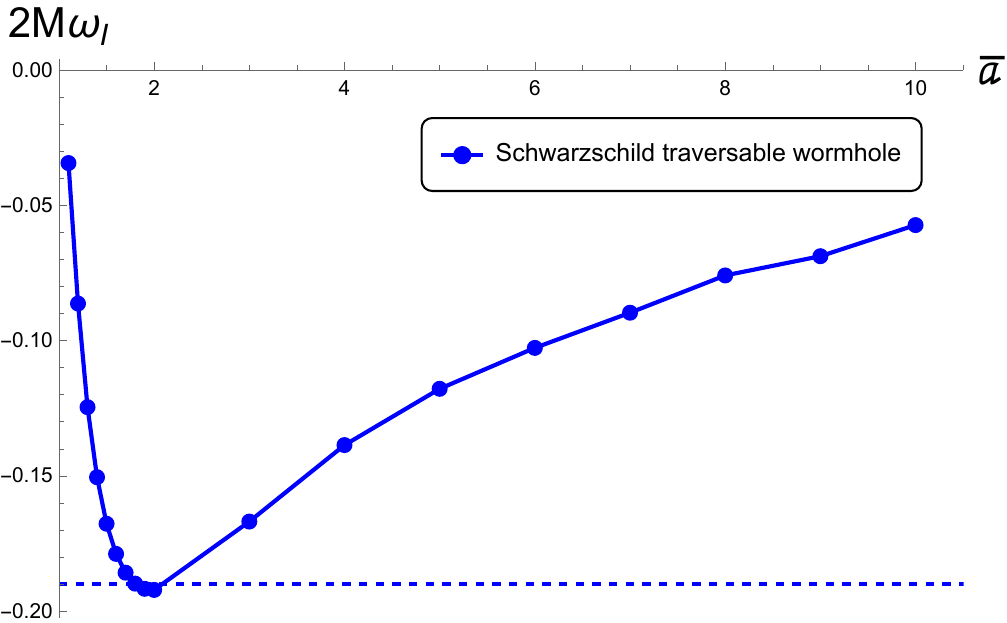}
    \end{minipage}
}
\caption{The two plots show how the real and imaginary parts of QNMs for Schwarzschild traversable wormholes vary with the parameter $\bar{a}$ under the electromagnetic field perturbation with $l=2$ .  
The solid curves are the corresponding variation curves, where the dots represent relevant data given by Table~\ref{Tab:QNM_EM}.
The dotted lines are constant reference lines parallel to the $\bar{a}$ axis, and their fixed vertical coordinate values correspond to the real and imaginary parts of QNMs of the Schwarzschild black hole.
These fixed values are used to compare with the vertical coordinate values of the solid curves.}
\label{fig:QNMEM-1}
\end{figure}

\begin{figure}[htbp]
\centering

    \includegraphics[width=0.5\linewidth]{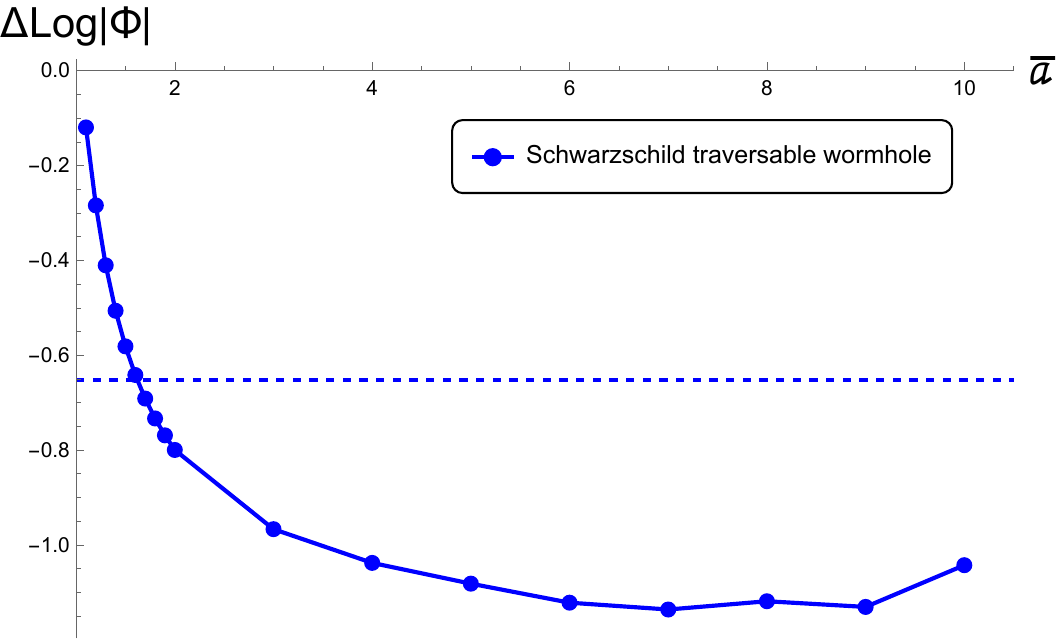}

\caption{This figure shows how the difference of $\log|\Phi|$ between adjacent peaks,  $\Delta\log|\Phi|$, for Schwarzschild traversable wormholes varies with the parameter $\bar{a}$ under the electromagnetic field perturbation with $l=2$. 
The solid curve is the corresponding variation curve, where the dots represent relevant data given by Table~\ref{Tab:QNM_EM}.
The dotted line is a constant reference line parallel to the $\bar{a}$ axis, and its fixed vertical coordinate value corresponds to  $\Delta\log|\Phi|$ of the Schwarzschild black hole.
This fixed value is used to compare with the vertical coordinate values of the solid curve.}
\label{fig:QNMEM-2}
\end{figure}

Next, we present, via numerical calculations, the plots of $2M\omega_{\text{R}}$ and $2M\omega_{\text{I}}$ with respect to $\bar{a}$ in Fig.~\ref{fig:QNMEM-1-a} and Fig.~\ref{fig:QNMEM-1-b}, respectively, where the dots denote the relevant data given in Table~\ref{Tab:QNM_EM}.
By using Eq.~\eqref{eq:Delta-Log} and Fig.~\ref{fig:QNMEM-1}, we obtain the plot of $\Delta\log|\Phi|$ with respect to $\bar{a}$ as shown in Fig.~\ref{fig:QNMEM-2}.
According to the QNM of Schwarzschild black holes, $2M\omega=0.915464 - 0.190100 i$ (see the second row in Table \ref{Tab:QNM_EM}), and Eq.~\eqref{eq:Delta-Log}, we get
\begin{equation}\label{eq:Delta_Phi-SCHEM}
     \Delta\log|\Phi|=\frac{- 0.190100}{0.915464}\pi\approx-0.652365.
\end{equation}
If the difference of $\log|\Phi|$ between adjacent peaks in the damping oscillation of electromagnetic field perturbations deviates significantly from Eq.~\eqref{eq:Delta_Phi-SCHEM}, we can infer that the spacetime corresponds to a Schwarzschild traversable wormhole rather than a Schwarzschild black hole.
Moreover, using the plot of $\log|\Phi|-t$, we can also determine the parameters $M$ and $a$ with the help of Figs.~\ref{fig:QNMEM-1-a} and \ref{fig:QNMEM-2}, as well as the scenario for fixing parameters introduced in Sec.~\ref{sec:method}.

\section{Axial gravitational field perturbation}\label{sec:Gravity}

\subsection{Echo waveform}\label{sec:GR_Echo}

\begin{figure}[htbp]
	\centering
	\subfigure[$\bar{a}=1.000001$]{
		\begin{minipage}[t]{0.4\linewidth}
			\centering
			\includegraphics[width=1\linewidth]{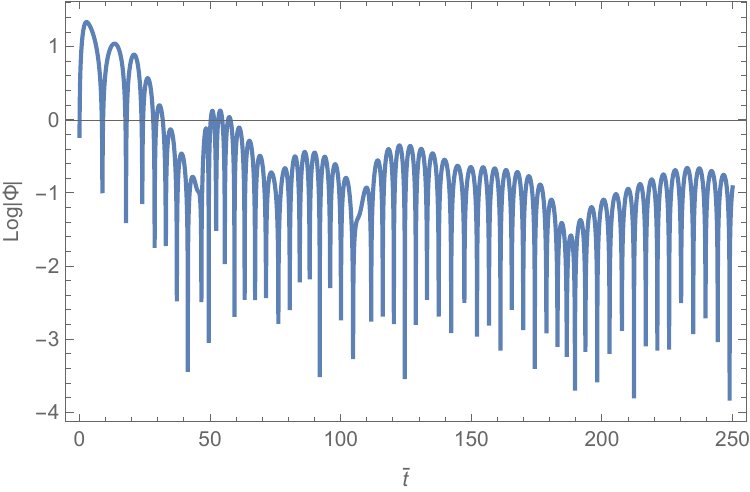}
		\end{minipage}
	}
	\subfigure[$\bar{a}=1.000001$]{
		\begin{minipage}[t]{0.4\linewidth}
			\centering
			\includegraphics[width=1\linewidth]{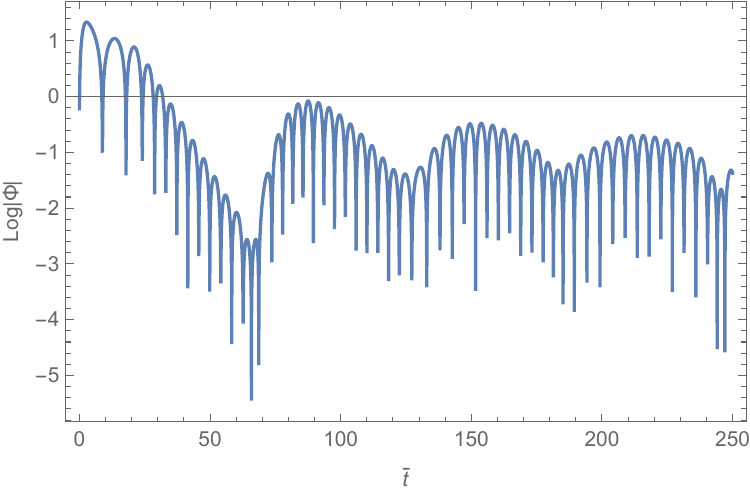}
		\end{minipage}
	}
	\vspace{-3mm}
	\subfigure[$\bar{a}=1.00001$]{
		\begin{minipage}[t]{0.4\linewidth}
			\centering
			\includegraphics[width=1\linewidth]{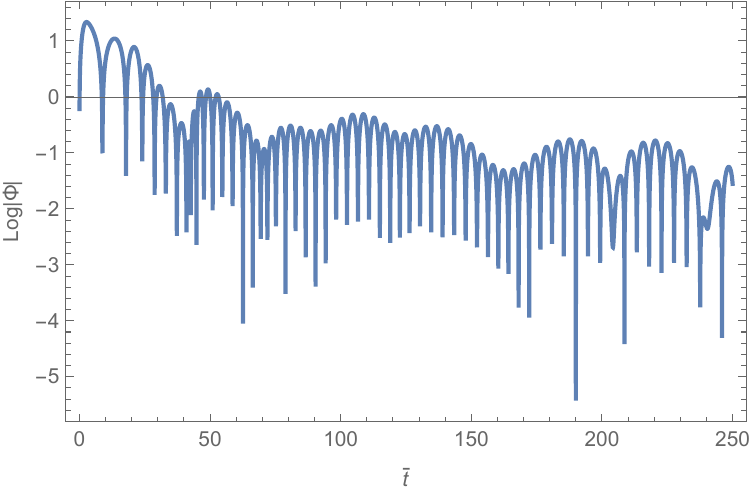}
		\end{minipage}
	}
	\subfigure[$\bar{a}=1.00001$]{
		\begin{minipage}[t]{0.4\linewidth}
			\centering
			\includegraphics[width=1\linewidth]{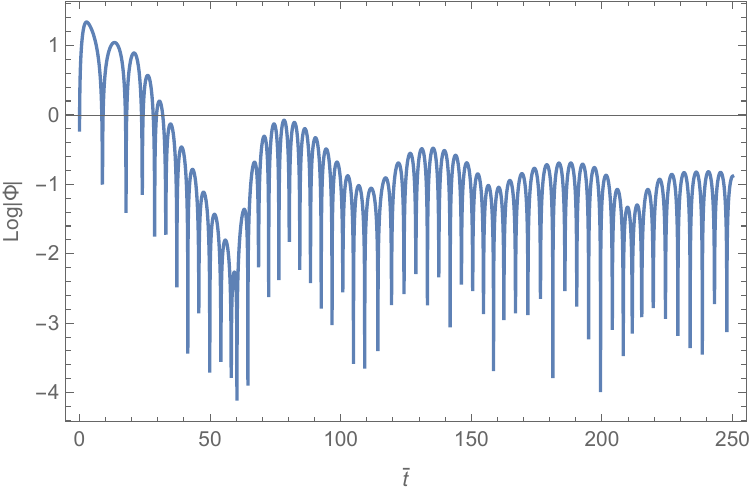}
		\end{minipage}
	}
	\vspace{-3mm}
	\subfigure[$\bar{a}=1.0001$]{
		\begin{minipage}[t]{0.4\linewidth}
			\centering
			\includegraphics[width=1\linewidth]{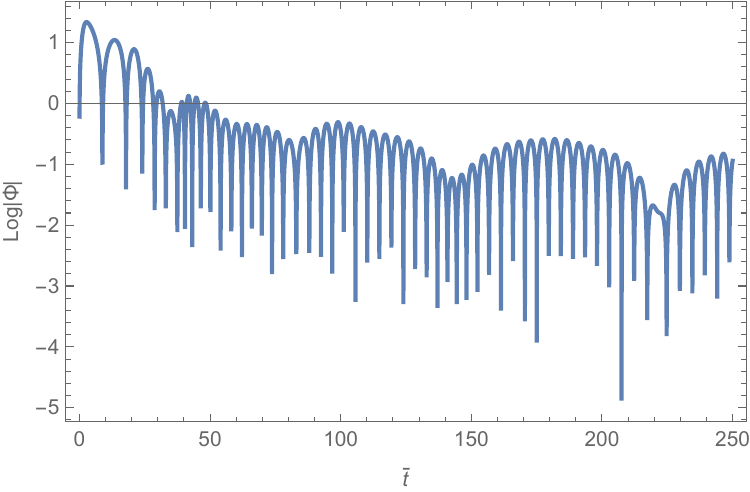}
		\end{minipage}
	}
	\subfigure[$\bar{a}=1.0001$]{
		\begin{minipage}[t]{0.4\linewidth}
			\centering
			\includegraphics[width=1\linewidth]{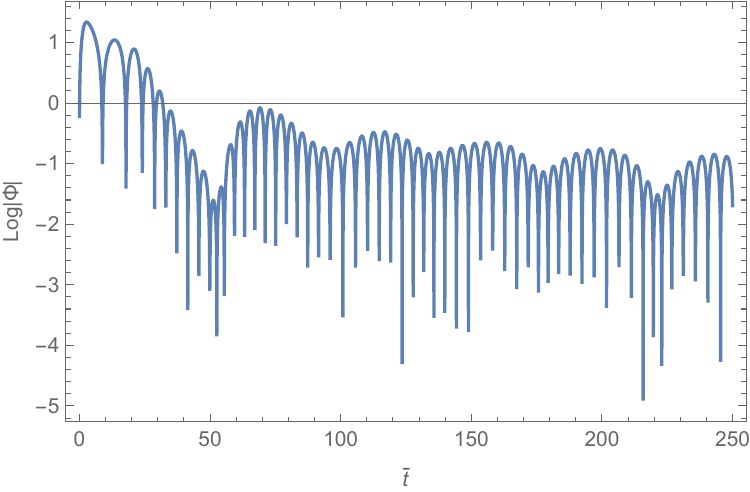}
		\end{minipage}
	}
	\vspace{-3mm}
	\subfigure[$\bar{a}=1.001$]{
		\begin{minipage}[t]{0.4\linewidth}
			\centering
			\includegraphics[width=1\linewidth]{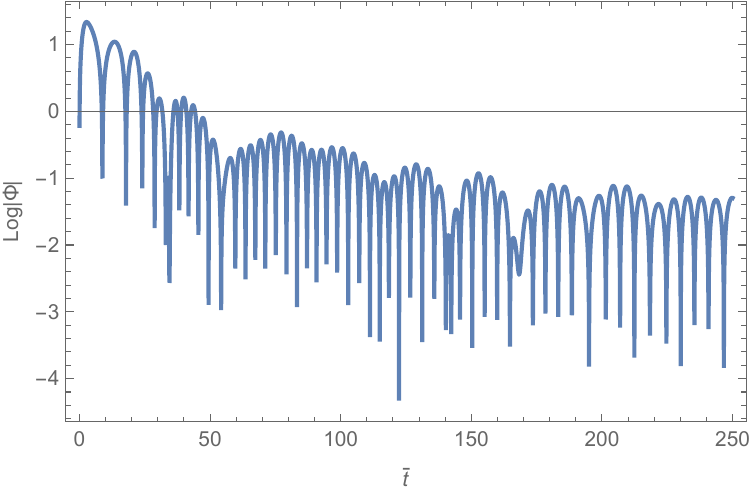}
		\end{minipage}
	}
	\subfigure[$\bar{a}=1.001$]{
		\begin{minipage}[t]{0.4\linewidth}
			\centering
			\includegraphics[width=1\linewidth]{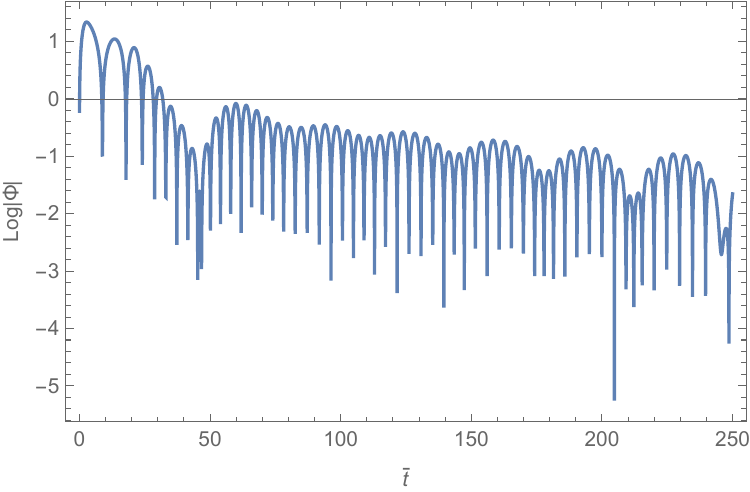}
		\end{minipage}
	}
	\caption{The echo waveforms around Schwarzschild traversable wormholes for different values of the parameter $\bar{a}$ under the axial gravitational field perturbation with $l=2$. The left diagrams show how the function $\log|\Phi|$ varies with the time $\bar{t}$ when the matter of the throat is considered, and the right ones show how the function $\log|\Phi|$ varies with the time $\bar{t}$ when the matter of the throat is ignored, i.e., $C_{s=2}(\bar{a})=0$.}
	\label{fig:gravity-echo}
\end{figure}

The echo waveform under the axial gravitational field perturbation also appears in the range where $\bar{a}>1$ and $\bar{a}-1\ll 1$.
In Fig.~\ref{fig:gravity-echo}, we present the echo waveforms with and without the influence of matter at the throat of the wormhole, respectively.
It can be seen that the axial gravitational perturbation echo waveform without the influence of matter at the throat is similar to the echo waveforms of scalar field perturbation and electromagnetic field perturbation.
However, the influence of the matter at the throat must be taken into account since the coefficient of the delta function term $C_{s=2}(\bar{a})$ cannot be ignored.
At this point, the main feature of the echo waveform of the axial gravitational field perturbation is that the occurrence time of its first echo signal is earlier than that of the first echo signal in the other two perturbation scenarios, and its peak is also higher.
The reason is that the first echo signal in the case of axial gravitational perturbations is the reflection of the incident wave by the $\delta$-function potential barrier at the throat of the wormhole, so it occurs earlier in time; while the first echo signal in the other two perturbation scenarios is the result of the incident wave being reflected by the potential barrier of the opposite universe.

\subsection{Damping oscillation}\label{sec:GR-Damping}
As $\bar{a}$ increases, the waveform of the axial gravitational field perturbation enters a damping oscillation. 
In Table~\ref{Tab:QNM_G_Odd-delta}, we present the QNMs of the waveform for different values of $\bar{a}$. 
For comparison, we also provide the QNMs without the influence of matter at the throat in Table~\ref{Tab:QNM_G_Odd}.
By comparing Table~\ref{Tab:QNM_G_Odd-delta} with Table~\ref{Tab:QNM_G_Odd}, we find that the presence of matter at the throat of the wormhole increases the oscillation frequency of the axial perturbation waveform and accelerates the decay of the waveform, which is consistent with the role played by the matter at the throat of the wormhole in the scalar field perturbation.
\begin{table}[!ht]
    \centering
    \caption{This table shows the waveform types and QNMs under the axial gravitational field perturbation with $l=2$. Here, the case of $\bar{a}\leq 1$  corresponds to Schwarzschild black holes, the case of $1.000001\leq\bar{a}\leq 1.001 $ corresponds to Schwarzschild traversable wormholes with obvious echoes, and the case of  $1.1\leq\bar{a}\leq 10.0 $ corresponds to Schwarzschild traversable wormholes with only damping oscillations.}
 \begin{tabular}{cccc}\hline
  
    $\bar{a}$          & waveform/QNM ($2M\omega$) & $\bar{a}$          & waveform/QNM ($2M\omega$)\\ \hline
  
    $\leq 1$           & 0.747450 - 0.178107 i &$1.7$              & 0.783574 - 0.264669 i\\ \hline
    $1.000001$   & Echo                        &$1.8$              & 0.759802 - 0.271159 i\\ \hline
    $1.00001$     & Echo                       &$1.9$              & 0.736425 - 0.274804 i\\ \hline
    $1.0001$       & Echo                      &$2.0$              & 0.713806 - 0.276322 i \\ \hline
    $1.001$          & Echo                    &$3.0$              & 0.535648 - 0.247569 i\\ \hline
    $1.06$             & 0.782781 - 0.039926 i (Weak echo)& $4.0$  & 0.423559 - 0.208944 i\\ \hline
    $1.1$              & 0.825691 - 0.072068 i &$5.0$              & 0.349308 - 0.178177 i\\ \hline
    $1.2$              & 0.862155 - 0.137343 i &$6.0$              & 0.296877 - 0.154676 i\\ \hline
    $1.3$              & 0.862515 - 0.183560 i &$7.0$              & 0.257297 - 0.136713 i\\ \hline
    $1.4$              & 0.849047 - 0.216146 i &$8.0$              & 0.228097 - 0.121663 i\\ \hline
    $1.5$              & 0.831069 - 0.238674 i &$9.0$              & 0.200610 - 0.093740 i\\ \hline
    $1.6$              & 0.807071 - 0.254348 i &$10.0$             & 0.171488 - 0.062760 i\\ \hline
  \end{tabular}
    \label{Tab:QNM_G_Odd-delta}
\end{table}

\begin{table}[!ht]
    \centering
    \caption{This table shows the waveform types and QNMs under the axial gravitational field perturbation with $l=2$ when the influence of the matter at the wormhole throat is ignored, i.e. $C_{s=2}(\bar{a})=0$. Here, the case of $\bar{a}\leq 1$  corresponds to Schwarzschild black holes, the case of $1.000001\leq\bar{a}\leq 1.001 $ corresponds to Schwarzschild traversable wormholes with obvious echoes, and the case of  $1.1\leq\bar{a}\leq 10.0 $ corresponds to Schwarzschild traversable wormholes  with only damping oscillations.}
  \begin{tabular}{cccc}\hline
  
    $\bar{a}$          & waveform/QNM ($2M\omega$) & $\bar{a}$          & waveform/QNM ($2M\omega$)\\ \hline
  
    $\leq 1$           & 0.747450 - 0.178107 i &$1.7$              & 0.727195 - 0.155295 i\\ \hline
    $1.000001$   & Echo                            &$1.8$              & 0.706166  - 0.160560 i\\ \hline
    $1.00001$     & Echo                             &$1.9$              & 0.685190 - 0.163871 i\\ \hline
    $1.0001$       & Echo                             &$2.0$              & 0.664565 - 0.165691 i \\ \hline
    $1.001$          & Echo                             &$3.0$              & 0.498185 - 0.152153 i\\ \hline
    $1.01$            & 0.741595 - 0.020092 i (Weak echo)& $4.0$ & 0.394052 - 0.129780  i\\ \hline
    $1.1$              & 0.725765 - 0.027264 i  &$5.0$              & 0.324647 - 0.111235 i\\ \hline
    $1.2$              & 0.777911 - 0.066573 i &$6.0$              & 0.275629 - 0.096729 i\\ \hline
    $1.3$              & 0.787784 - 0.097540 i &$7.0$              & 0.239082 - 0.085672 i\\ \hline
    $1.4$              & 0.780846  - 0.120006 i &$8.0$              & 0.211504 - 0.076267 i\\ \hline
    $1.5$              & 0.766049 - 0.136103 i &$9.0$              & 0.189583 - 0.069135 i\\ \hline
    $1.6$              & 0.747529 - 0.147468 i &$10.0$            &  0.171488 - 0.062760 i\\ \hline
  \end{tabular}
    \label{Tab:QNM_G_Odd}
\end{table}

\begin{figure}[htbp]
\centering
\subfigure[Real part of QNMs]{\label{fig:GRQNM-1-a}
    \begin{minipage}[t]{0.45\linewidth}
    \centering
    \includegraphics[width=1\linewidth]{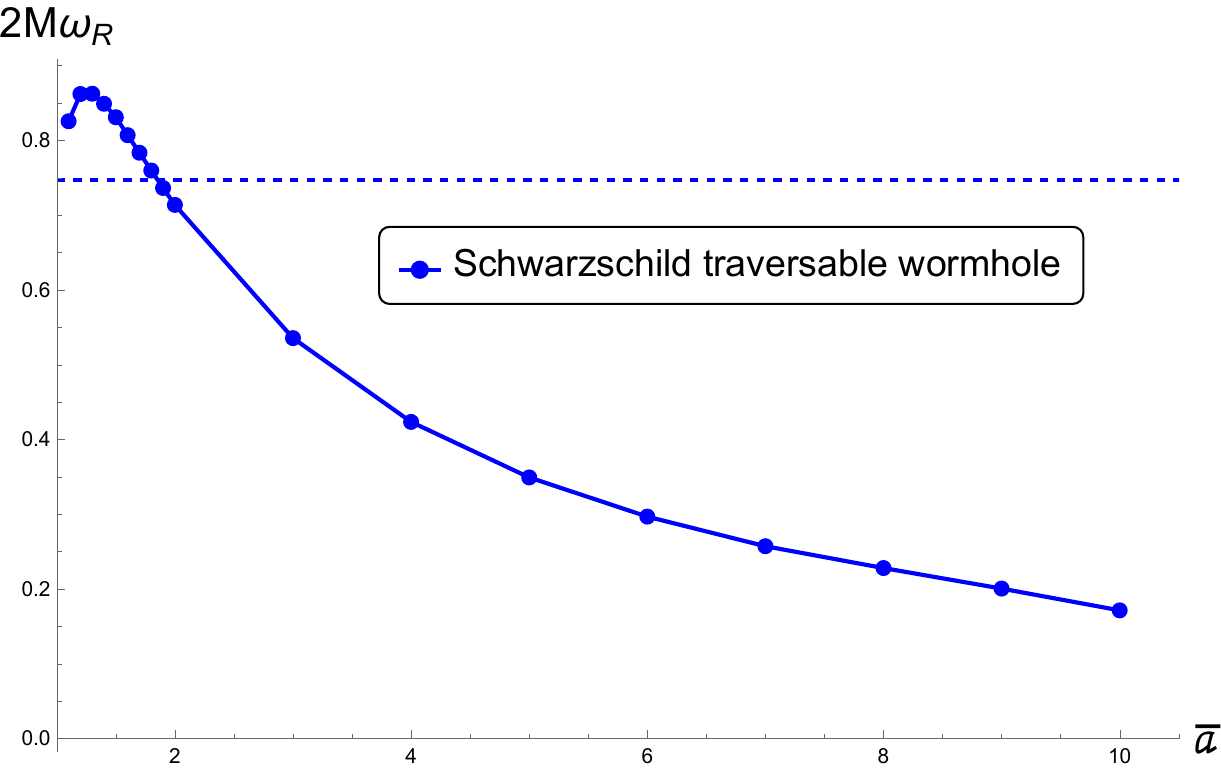}
    \end{minipage}
}
\subfigure[Imaginary part of QNMs]{\label{fig:GRQNM-1-b}
    \begin{minipage}[t]{0.45\linewidth}
    \centering
    \includegraphics[width=1\linewidth]{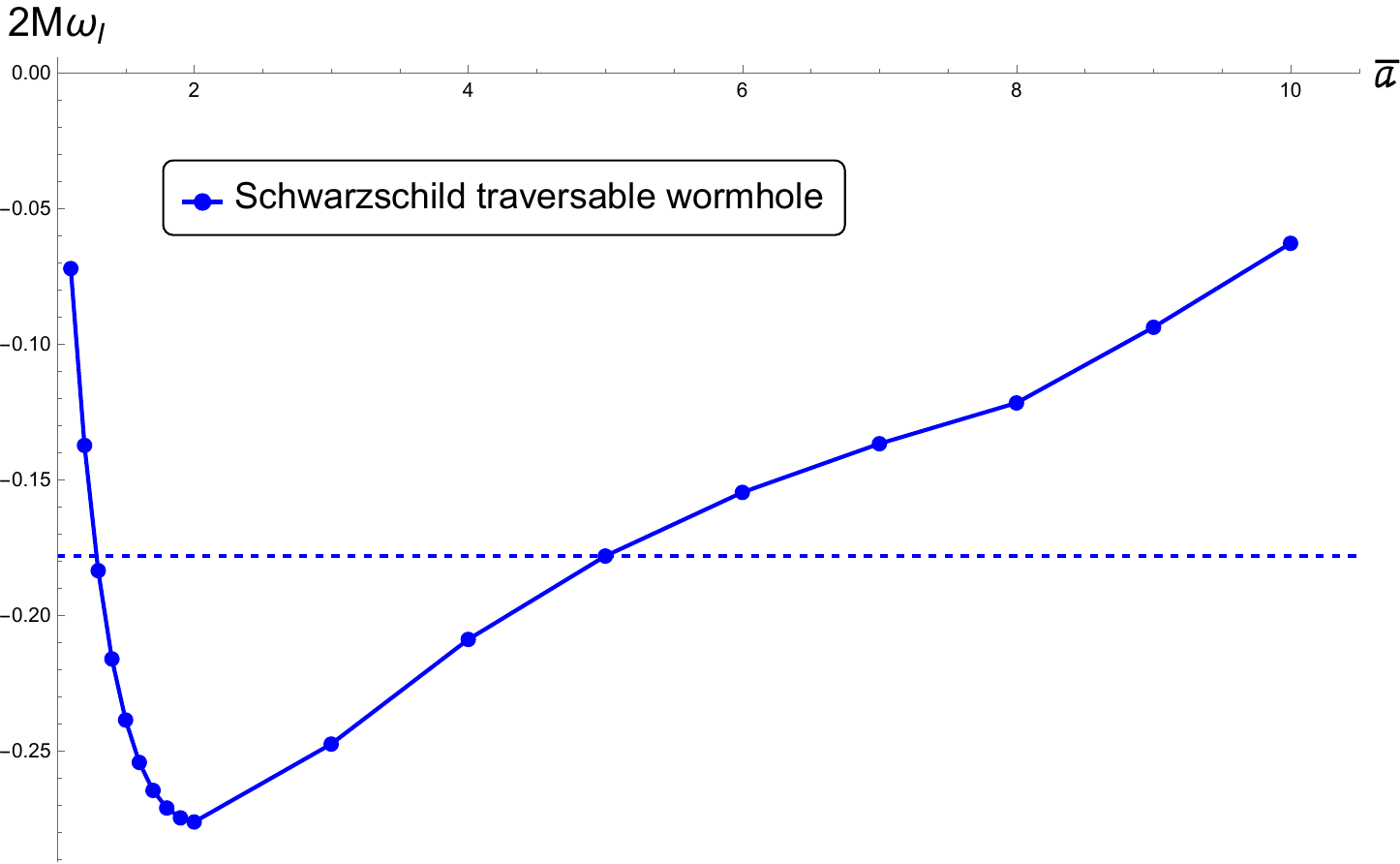}
    \end{minipage}
}
\caption{The figure shows how the real and imaginary parts of QNMs for Schwarzschild traversable wormholes vary with the parameter $\bar{a}$ under the axial gravitational field perturbation with $l=2$. 
The solid curves are the corresponding variation curves, where the dots represent relevant data given by Table ~\ref{Tab:QNM_G_Odd}.
The dotted lines are constant reference lines parallel to the $\bar{a}$ axis, and their fixed vertical coordinate values correspond to the real and imaginary parts of QNMs of the Schwarzschild black hole.
These fixed values are used to compare with the vertical coordinate values of the solid curves.}
\label{fig:GRQNM-1}
\end{figure}

\begin{figure}[htbp]
\centering

    \includegraphics[width=0.5\linewidth]{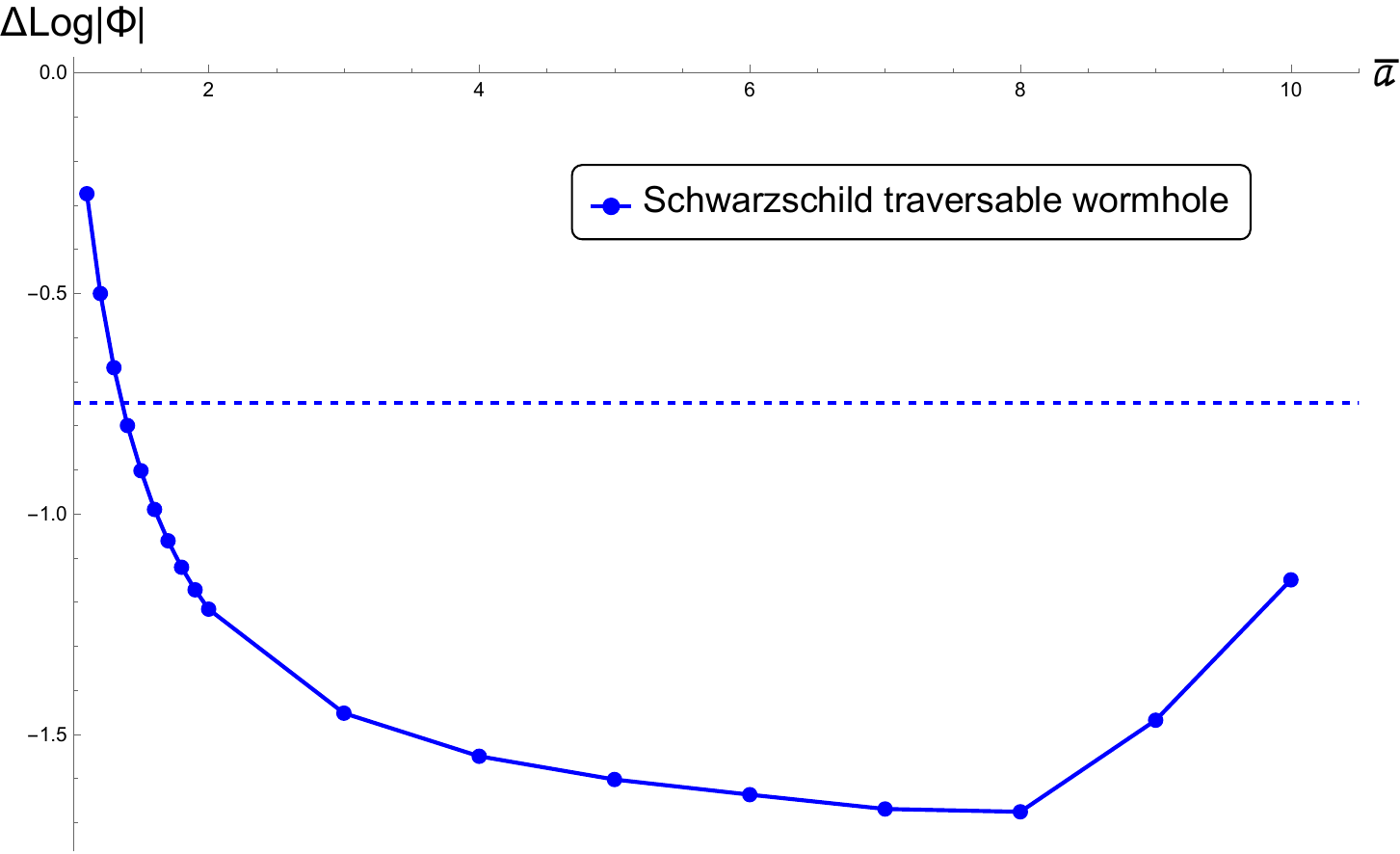}

\caption{This figure shows how the difference of $\log|\Phi|$ between adjacent peaks,  $\Delta\log|\Phi|$, for Schwarzschild traversable wormholes varies with parameter $\bar{a}$ under the axial gravitational field perturbation with $l=2$ in the case of Schwarzschild traversable wormholes. 
The solid curve is the corresponding variation curve, where the dots represent relevant data given by Table ~\ref{Tab:QNM_G_Odd}.
The dotted line is a constant reference line parallel to the $\bar{a}$ axis, and its fixed vertical coordinate value corresponds to  $\Delta\log|\Phi|$ of the Schwarzschild black hole.
This fixed value is used to compare with the vertical coordinate values of the solid curve.}
\label{fig:GRQNM-2}
\end{figure}

Finally, we present the plots of $2M\omega_{\text{R}}$ and $2M\omega_{\text{I}}$ varying with $\bar{a}$ in Figs.~\eqref{fig:GRQNM-1-a} and \ref{fig:GRQNM-1-b}, respectively, where the blue dots depend on the relevant data in Table~\ref{Tab:QNM_G_Odd-delta}.
According to Fig.~\ref{fig:GRQNM-1} and Eq.~\eqref{eq:Delta-Log}, we plot $\Delta\log|\Phi|$  with respect to $\bar{a}$ in Fig.~\ref{fig:GRQNM-2}. For Schwarzschild black holes, we compute $\Delta\log|\Phi|$ by using the QNM of axial gravitational field perturbations, $2M\omega=0.747450 - 0.178107 i$, 
\begin{equation}\label{eq:Delta_Phi-SCHGO}
     \Delta\log|\Phi|=\frac{- 0.178107}{0.747450}\pi\approx-0.748598.
\end{equation}
Therefore, we may distinguish a Schwarzschild traversable wormhole from a Schwarzschild black hole by measuring such a quantity.
In addition, we can determine the parameters $M$ and $a$ in terms of the plot of $\log|\Phi|-t$ by observations, together with Figs.~\ref{fig:GRQNM-1} and \ref{fig:GRQNM-2}, when following the scenario introduced in Sec.~\ref{sec:method}.
In particular, we can determine the range of parameters more accurately by jointly considering various field perturbations, such as the scalar field, the electromagnetic field, and the axial gravitational field perturbations.

\newpage
\section{Conclusion}\label{sec:Con}
We investigate the properties of Schwarzschild traversable wormholes and find their crucial factor --- the position of throats, where such a model is primarily obtained by applying the ``cut-and-paste” technique in the Schwarzschild spacetime.
Meanwhile, we fully consider the influence of the matter term at the throat of the wormhole on the scalar field perturbation, the electromagnetic field perturbation, and the axial gravitational field perturbation in spacetime, and present the corresponding perturbation equations.
Moreover, we numerically calculate the waveforms and QNMs of the scalar field perturbation, the electromagnetic field perturbation, and the axial gravitational field perturbation for the model.
By comparing the properties of Schwarzschild traversable wormholes with those of Schwarzschild black holes, we obtain the following unique features for the former:
\begin{itemize}
\item The matter at the throat of a wormhole affects the propagation of scalar field perturbations and axial gravitational field perturbations, but dose not affect the propagation of electromagnetic field perturbations.
\item 
The waveform of the three field perturbations exhibits the echo waveform when the throat parameter $a$ and the mass parameter $M$ of a Schwarzschild traversable wormhole satisfy the condition $\frac{a}{2M}\ll 1$. 
However, the echo waveform does not appear in a Schwarzschild black hole.
\item The matter at the throat of the wormhole can be ignored for the echo waveform of the scalar field, leading to similar patterns in the echo waveforms of the scalar field and the electromagnetic field. However, the matter at the throat of the wormhole has a significant impact on the echo waveform of the axial gravitational field, mainly manifested in the earlier occurrence time and larger amplitude of the first echo signal.
\item In the damping oscillation waveform, the matter at the throat affects the QNMs of the scalar field perturbation waveform and the axial gravitational field perturbation waveform. It manifests mainly as a higher oscillation frequency and a faster decay rate compared to the case in which the influence of matter is not considered.
\item In a damping oscillation waveform depicted in the plot of $\log|\Phi|-t$, the difference between adjacent peaks varies with the change of $\frac{a}{2M}$ in a Schwarzschild traversable wormhole, but remains constant in a Schwarzschild black hole.

\end{itemize}

Based on these unique features, we summarize a scenario for estimating the parameters $M$ and $a$ from the plot of $\log|\Phi|-t$ and the numerical results of QNMs.
Although it is only a theoretical concept at present, a Schwarzschild traversable wormhole is expected to be observed in future gravitational wave experiments through the above-mentioned features.
If so, we shall gain a deeper understanding and more knowledge of the universe.

In the research on the quantization of the event horizon of black holes, some studies~\cite{Cardoso:2019apo,Li:2019kwa} have proposed introducing some special potentials near the event horizon to simulate the quantization of the event horizon. 
However, the material sources of these special potentials remain unclear.
We find that the matter at the throat of a wormhole generates a $\delta$-function potential at the throat in the scalar field perturbation and axial gravitational field perturbation equations. 
This might provide an explanation for the material sources of the quantized event horizon, thereby enhancing our understanding of the quantization of the event horizon.


\section*{Acknowledgments}
Y-GM would like to thank Emmanuele Battista, Stenfan Fredenhagen, and Harold Steinacker for the warm hospitality during his stay at University of Vienna.  
This work was supported in part by the National Natural Science Foundation of China under Grant No.\ 12175108.


\section*{Appendix}
\begin{appendices}
\section{Derivation of perturbation equations  \label{appendix:A}}
\setcounter{equation}{0}
\renewcommand\theequation{A.\arabic{equation}}
Here we use the linear perturbation theory~\cite{Berti:2009kk,Maggio:2020jml,Rahman:2021kwb} to derive the perturbation equations in Schwarzschild traversable wormholes.
For the scalar and electromagnetic field perturbations, we can derive the perturbation equations directly from their equations of motion since there is no corresponding background field in Schwarzschild traversable wormholes.
In the case of scalar field perturbations, we adopt the Klein-Gordon equation,
\begin{equation}\label{Aeq:KG}
    \nabla^\mu\nabla_\mu\Psi=0.
\end{equation}
Because the metric is independent of time and has the spherical symmetry, we decompose $\Psi(t,r_*,\theta,\phi)$ as
\begin{equation}
    \Psi(t,r_*,\theta,\phi)=e^{-i\omega t}\sum_{l,m}\frac{\Phi_{s=0}(r_*)}{\chi(r_*)}Y_{lm}(\theta,\phi),
\end{equation}
where $Y_{lm}(\theta,\phi)$ denotes the spherical harmonics, and $r_*$ is the tortoise coordinate defined by Eq.~\eqref{wh-tc} for traversable wormholes.
Through expanding Eq.~\eqref{Aeq:KG} under the metric Eq.~\eqref{tc-metric}, we can write its radial part in the form of a Schrödinger-like equation,
\begin{equation}\label{Aeq:schlike}
    \frac{\dif^2\Phi_{s=0}}{\dif r_*^2}+(\omega^2-V_{s=0})\Phi_{s=0}=0,
\end{equation}
where 
\begin{equation}\label{Aeq:Vs0}
    V_{s=0}=\left(1-\frac{2M}{\chi}\right)\left[\frac{l(l+1)}{\chi^2}+\frac{\chi''}{\chi-2M}\right].
\end{equation}
According to Eq.~\eqref{chi''}, we then give the effective potential outside the throat of Schwarzschild traversable wormholes,
\begin{equation}\label{whspot}
    V_{s=0}(\chi\neq a)=\left(1-\frac{2M}{\chi}\right)\left[\frac{l(l+1)}{\chi^2}+\frac{2M}{\chi^3}\right],
\end{equation}
and the potential at the throat,
\begin{equation}\label{whspat}
    V_{s=0}(\chi=a)=\left(1-\frac{2M}{a}\right)\left[\frac{l(l+1)}{a^2}+\frac{2}{a}\delta(0)\right].
\end{equation}

In the case of electromagnetic field perturbations, we adopt Maxwell's equations,
\begin{equation}\label{Aeq:Max}
    \nabla_\nu F^{\mu\nu}=0,  \qquad F_{\mu\nu}=\partial_\mu A_\nu-\partial_\nu A_\mu,
\end{equation}
where $A_\mu$ is the vector potential.
Here we can expand~\cite{Cardoso:2001bb} $A_\mu$ in the four-dimensional vector formulation,
\begin{equation}\label{Aeq:Amu}
    A_\mu(t,\chi,\theta,\phi)=\sum_{l,m}\left(
\begin{bmatrix}
0\\ 0\\ e^{-i\omega t}\frac{a^{lm}(\chi)}{\sin\theta}\partial_\phi Y_{lm}\\-e^{-i\omega t}a^{lm}(\chi)\sin\theta\partial_\theta Y_{lm}
\end{bmatrix}+
\begin{bmatrix}
e^{-i\omega t}f^{lm}(\chi)Y_{lm}\\ e^{-i\omega t}h^{lm}(\chi)Y_{lm}\\e^{-i\omega t} k^{lm}(\chi)\partial_\theta Y_{lm}\\e^{-i\omega t} k^{lm}(\chi)\partial_\phi Y_{lm}
\end{bmatrix}
    \right),
\end{equation}
where the first term has the odd parity $(-1)^{l+1}$ and the second one has the even parity $(-1)^l$ under the transformation $\theta\rightarrow -\theta$,
and $a^{lm}$, $f^{lm}$, and $k^{lm}$ are functions of  $\chi$. 
By substituting Eq.~\eqref{Aeq:Amu} into Eq.~\eqref{Aeq:Max}, we can get an equation similar to Eq.~\eqref{Aeq:schlike}, where the corresponding radial wave function $\Phi_{s=1}$ takes the form, 
\begin{equation}\label{AeqPhis1}
    \Phi_{s=1}=\begin{cases}
a^{lm}, \qquad \qquad\qquad\qquad \quad    \text{for odd parity,}  \\
\frac{\chi^3}{\chi-2M}(\frac{\dif f^{lm}}{\dif r_*}+i\omega h^{lm}), \quad\;\,  \text{for even parity,}  
    \end{cases}
\end{equation}
and the potential $V_{s=1}$ is
\begin{equation}\label{Aeq:Vs1}
    V_{s=1}=\left(1-\frac{2M}{\chi}\right)\frac{l(l+1)}{\chi^2}.
\end{equation}

For gravitational perturbations, we introduce a perturbation $h_{\mu\nu}$ on the background spacetime $g_{\mu\nu}$ (Eq.~\eqref{eq:metricSch}) and then write the perturbed spacetime, 
\begin{equation}
    \hat{g}_{\mu\nu}=g_{\mu\nu}+h_{\mu\nu}.
\end{equation}
According to the parity behavior under the transformation $\theta\rightarrow-\theta$, the perturbation $h_{\mu\nu}$ can be divided~\cite{regge1957stability,zerilli1970effective} into two classes: the axial (odd parity) perturbation with parity $(-1)^{l+1}$, and the polar (even parity) perturbation with parity $(-1)^l$. 
In the Regge-Wheeler gauge~\cite{regge1957stability}, the axial perturbation can be written as
\begin{equation}
    h^{-}_{\mu\nu}=\begin{bmatrix}
        0 & 0 & 0 & h_0(r_*)\\
        0 & 0 & 0 & h_1(r_*)\\
        0 & 0 & 0 & 0 \\
        h_0(r_*) & h_1(r_*) & 0 & 0 \\
    \end{bmatrix}
    e^{-i\omega t}\left(\sin\theta\frac{\partial}{\partial \theta}\right)Y_{l0}(\theta).
\end{equation}
The variation of the Ricci tensor before and after perturbation is
\begin{equation}
    \label{delta-Ricci}
    \delta R_{\mu\nu}=\hat{R}_{\mu\nu}-R_{\mu\nu}\approx\nabla_\rho\delta\Gamma^\rho_{\nu\mu}-\nabla_\nu\delta\Gamma^\rho_{\rho\mu},
\end{equation}
where only the linear term of the perturbation is retained and $\hat{R}_{\mu\nu}$ is the Ricci tensor of the perturbed spacetime, and
\begin{equation}
    \label{delta-Gamma}
    \delta\Gamma^\alpha_{\beta\gamma}=\frac{1}{2}g^{\alpha\nu}\left(\nabla_\gamma h_{\beta\nu}+\nabla_\beta h_{\gamma\nu}-\nabla_\nu h_{\beta\gamma}\right).
\end{equation}
Based on the Einstein field equations Eq.~\eqref{Einstein-eq}, we can obtain the linear terms of the perturbation,
\begin{equation}
    \label{delta-Einstein}
    \delta R_{\mu\nu}-\frac{1}{2}R\cdot h_{\mu\nu}-\frac{1}{2}\delta R\cdot g_{\mu\nu}=8\pi\delta T_{\mu\nu}.
\end{equation}
For the axial perturbation, the components that can be used to derive the perturbation equation are
\begin{equation}
    \label{deltaR13}
    \delta R_{13}-\frac{1}{2}R\cdot h_{13}=0,
\end{equation}
and 
\begin{equation}
    \label{deltaR23}
    \delta R_{23}=0.
\end{equation}
By solving Eqs.~\eqref{deltaR13} and ~\eqref{deltaR23}, we get two equations,
\begin{equation}
    \label{R23=0}
    i\omega h_0+\frac{\dif h_1(r_*)}{\dif r_*}=0,
\end{equation}
and 
\begin{equation}
    \label{R13=0}
    \begin{split}
        &-i\omega\chi^3\frac{\dif h_0(r_*)}{\dif r_*}+2i\omega\chi^2\frac{\dif \chi}{\dif r_*}h_0(r_*)\\
        &+\left[\omega^2\chi^3+2\chi\left(\frac{\dif \chi}{\dif r_*}\right)^2+2r^2\frac{\dif^2\chi}{\dif r_*^2}+l(l+1)(2M-\chi)\right]h_1(r_*)-\frac{2M-\chi}{\chi^2}R\cdot h_1(r_*)=0.
    \end{split}
\end{equation}
With the transformation ${h_1(r_*)}={\chi}\Phi_{s=2}^-$, we obtain the axial perturbation equation,
\begin{equation}
    \label{odd-ppeq}
    \frac{\dif^2\Phi^-_{s=2}}{\dif r_*^2}+(\omega^2-V^-_{s=2})\Phi^-_{s=0}=0,
\end{equation}
where
\begin{equation}
V^-_{s=2}=\frac{\chi-2M}{\chi^3}l(l+1)-\frac{3}{\chi}\frac{\dif^2 \chi}{\dif r_*^2}+\frac{2M-\chi}{\chi}R.
\end{equation}
According to Eq.~\eqref{chi''}, we give the effective potential outside the throat of Schwarzschild traversable wormholes,
\begin{equation}\label{whgopot}
    V^-_{s=2}(\chi\neq a)=\left(1-\frac{2M}{\chi}\right)\left[\frac{l(l+1)}{\chi^2}-\frac{6M}{\chi^3}\right],
\end{equation}
and the potential at the throat,
\begin{equation}\label{whg0pat}
    V^-_{s=2}(\chi=a)=\frac{12M^2-8Ma}{a^4}+\frac{a-2M}{a^3}l(l+1)+\frac{2a-2M}{a^2}\delta(0).
\end{equation}

\end{appendices}

\begin{appendices}
\section{Derivation of difference equations  \label{appendix:B}}
\setcounter{equation}{0}
\renewcommand\theequation{B.\arabic{equation}}
 
We consider the differential equation,
\begin{equation}
    \label{eq:Peg-dl-B}
    \frac{\dif^2\Phi_s}{\dif \bar{r}_*^2}-\frac{\dif^2\Phi_s}{\dif \bar{t}^2}-\bar{V}_s\Phi_s-C_s(\bar{a})\delta(\bar{r}_*-\bar{r}_t)\Phi_s=0,
\end{equation}
where $\bar{V}_s$ represents the effective potential excluding the $\delta$-function, and $\bar{r}_t$ the position of the wormhole throat in the tortoise coordinate, that is, 
\begin{equation}
    \bar{r}_t=\bar{r}_*(\bar{\chi}=\bar{a})=\bar{a}+\ln(\bar{a}-1),
\end{equation}
and
\begin{equation}\label{csa-B}
    C_s(\bar{a})=\left\{
    \begin{aligned}
        & \frac{2(\bar{a}-1)}{\bar{a}^2},&s=0;\\
        & 0,&s=1 ;\\
        & \frac{2\bar{a}-1}{\bar{a}^2},&s=2.
    \end{aligned}
    \right.
\end{equation}
Integrating Eq.~\eqref{eq:Peg-dl-B} on both sides of the throat and taking the interval of integration to be zero,
\begin{equation}
    \lim_{\epsilon\rightarrow0}\int^{\bar{r}_t+\epsilon}_{\bar{r}_t-\epsilon}\left[\frac{\dif^2\Phi_s}{\dif \bar{r}_*^2}-\frac{\dif^2\Phi_s}{\dif \bar{t}^2}-\bar{V}_s\Phi_s-C_s(\bar{a})\delta(\bar{r}_*-\bar{r}_t)\Phi_s\right]\dif \bar{r}_*=0,
\end{equation}
we obtain
\begin{equation}\label{eq:discon}
    \left.\frac{\dif\Phi_s}{\dif\bar{r}_*}\right|_{\bar{r}_t+0^+}- \left.\frac{\dif\Phi_s}{\dif\bar{r}_*}\right|_{\bar{r}_t+0^-}=C_s(\bar{a})\Phi_s(\bar{r}_t),
\end{equation}
which indicates that the first-order derivative of $\Phi_s$ with respect to the tortoise coordinate undergoes a discontinuity at the throat due to the influence of the $\delta$-function.

Then, we start with discretizing the coordinate space $(\bar{t},\bar{r}_*)$, so that $(\bar{t},\bar{r}_*)$ becomes $(\bar{t}_0+j\Delta \bar{t},\bar{r}_{*0}+k\Delta \bar{r}_*)$, where $\bar{t}_0$ and $\bar{r}_{*0}$ are the coordinates of the grid origin, $j$ and $k$ are integers, and $\Delta \bar{t}$ and $\Delta \bar{r}_*$ are step sizes of coordinates.
For the sake of convenience, we write $(\bar{t}_0+j\Delta \bar{t},\bar{r}_{*0}+k\Delta \bar{r}_*)$ as $(j,k)$.
Meanwhile, we denote the values of $\bar{t}$ and $\bar{r}_*$ at the grid points $(j,k)$ as $\bar{t}_j$ and $\bar{r}_{*k}$.
When choosing the grid points, we adjust $\bar{t}_0$ and $\bar{r}_{*0}$ so that the throat of the wormhole does not fall on the grid points.
At the grid point $(j,k)$, Eq.~\eqref{eq:Peg-dl-B} can be written in the form of a difference equation,
\begin{equation}\label{eq:Peg_FD-B}
-\frac{\Phi_s(j+1,k)-2\Phi_s(j,k)+\Phi_s(j-1,k)}{\Delta \bar{t}^2}+\frac{\dif^2\Phi_s}{\dif \bar{r}_*^2}-\bar{V}_s(k)\Phi_s(j,k)=0,
\end{equation}
where the difference form of $\dif^2\Phi_s/d\bar{r}_*^2$ still needs to be further calculated.
Next, we try to get the difference form of $\dif^2\Phi_s/d\bar{r}_*^2$ with the grid points $(j,k-1)$, $(j,k)$ and $(j,k+1)$.

When the throat of the wormhole is outside the intervals of the grid points $(j,k-1)$, $(j,k)$ and $(j,k+1)$, the difference form of $\dif^2\Phi_s(j,k)/d\bar{r}_*^2$ is
\begin{equation}\label{eq:SOD-DIF}
    \frac{\dif^2\Phi_s(j,k)}{\dif \bar{r}_*^2}=\frac{\frac{\dif\Phi_s(j,k+\frac{1}{2})}{\dif\bar{r}_*}-\frac{\dif\Phi_s(j,k-\frac{1}{2})}{\dif\bar{r}_*}}{\Delta\bar{r}_*}=\frac{\Phi_s(j,k+1)-2\Phi_s(j,k)+\Phi_s(j,k-1)}{\Delta\bar{r}_*^2},
\end{equation}
where 
\begin{equation}
    \frac{\dif\Phi_s(j,k+\frac{1}{2})}{\dif\bar{r}_*}=\frac{\Phi_s(j,k+1)-\Phi_s(j,k)}{\Delta\bar{r}_*},\qquad \frac{\dif\Phi_s(j,k-\frac{1}{2})}{\dif\bar{r}_*}=\frac{\Phi_s(j,k)-\Phi_s(j,k-1)}{\Delta\bar{r}_*}.
\end{equation}

\begin{figure}[htbp]
\centering
\includegraphics[width=0.5\linewidth]{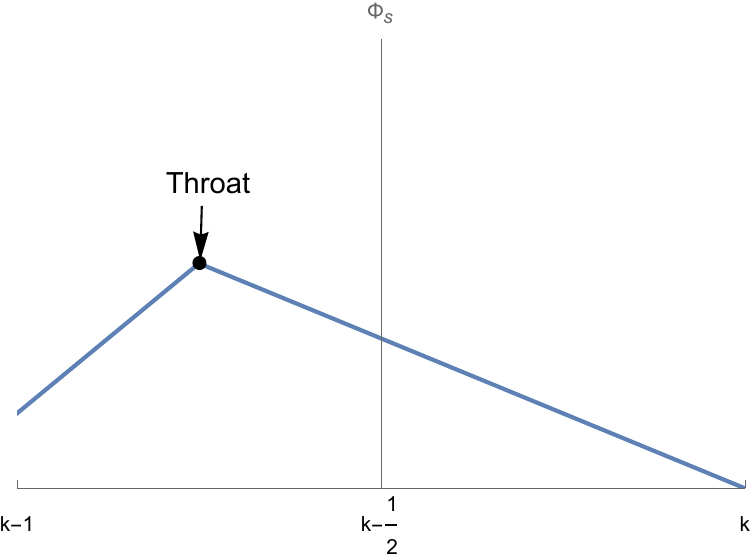}
\caption{The schematic diagram of $\Phi_s$ varying with the tortoise coordinate within the grid points $(j,k-1)$ and $(j,k)$. This diagram is plotted according to Eq.~\eqref{eq:discon}, and the wormhole throat $\bar{r}_t$ is within the range of $(\bar{r}_{*k-1},\bar{r}_{*k-\frac{1}{2}})$.}
\label{fig:FOD-1-1/2}
\end{figure}

When $\bar{r}_t\in(\bar{r}_{*k-1},\bar{r}_{*k-\frac{1}{2}})$, the discontinuity of $\dif\Phi_s/\dif\bar{r}_*$ at the throat directly affects the calculation of $\dif\Phi_s(j,k-\frac{1}{2})/\dif\bar{r}_*$.
In Fig.~\ref{fig:FOD-1-1/2}, we present the plots of $\Phi_s$ varying with the tortoise coordinate between the grids $(j, k-1)$ and $(j, k)$. 
Because the step size is taken to be small enough, we can assume that $\Phi_s$ varies linearly along the tortoise coordinate except at the throat.
Therefore, by using Fig.~\ref{fig:FOD-1-1/2} and Eq.~\eqref{eq:discon}, we obtain the equations,
\begin{equation}\label{eq:fod-1-12}
    \Phi_s(j,k)-\Phi_s(j,k-1)=\frac{\dif\Phi_s(j,k-1)}{\dif\bar{r}_*}(\bar{r}_t-\bar{r}_{*k-1})+\frac{\dif\Phi_s(j,k-\frac{1}{2})}{\dif\bar{r}_*}(\bar{r}_{*k}-\bar{r}_t)
\end{equation}
and
\begin{equation}\label{eq:fod-1-12-2}
    \frac{\dif\Phi_s(j,k-\frac{1}{2})}{\dif\bar{r}_*}=\frac{\dif\Phi_s(j,k-1)}{\dif\bar{r}_*}+C_s(\bar{a})\Phi_s(j,\bar{r}_t).
\end{equation}
After simplifying these two equations, we get
\begin{equation}
    \frac{\dif\Phi_s(j,k-\frac{1}{2})}{\dif\bar{r}_*}=\frac{\Phi_s(j,k)-\Phi_s(j,k-1)}{\Delta \bar{r}_*}+C_s(\bar{a})\Phi_s(j,\bar{r}_t)\frac{\bar{r}_t-\bar{r}_{*k-1}}{\Delta\bar{r}_*}.
\end{equation}
Finally, we can obtain the difference form of $\dif^2\Phi_s/d\bar{r}_*^2$ as
\begin{equation}\label{eq:SOD-1-12}
\begin{split}
    \frac{\dif^2\Phi_s(j,k)}{\dif \bar{r}_*^2}&=\frac{\frac{\dif\Phi_s(j,k+\frac{1}{2})}{\dif\bar{r}_*}-\frac{\dif\Phi_s(j,k-\frac{1}{2})}{\dif\bar{r}_*}}{\Delta\bar{r}_*}\\
    &=\frac{\Phi_s(j,k+1)-2\Phi_s(j,k)+\Phi_s(j,k-1)}{\Delta\bar{r}_*^2}-C_s(\bar{a})\Phi_s(j,\bar{r}_t)\frac{\bar{r}_t-\bar{r}_{*k-1}}{\Delta\bar{r}_*^2},\\
\end{split}
\end{equation}
where the specific form of $\Phi_s(j,\bar{r}_t)$ needs to be further determined.
According to Fig.~\ref{fig:FOD-1-1/2}, we expand Eq.~\eqref{eq:fod-1-12-2} in the following form,
\begin{equation}
    \frac{\Phi_s(j,\bar{r}_t)-\Phi(j,k-1)}{\bar{r}_t-\bar{r}_{*k-1}}+C_s(\bar{a})\Phi_s(j,\bar{r}_t)=\frac{\Phi_s(j,\bar{r}_t)-\Phi(j,k)}{\bar{r}_t-\bar{r}_{*k}},
\end{equation}
and thereby determine $\Phi_s(j,\bar{r}_t)$ to be
\begin{equation}\label{eq:phisthroat-1-12}
    \Phi_s(j,\bar{r}_t)=\frac{(\bar{r}_t-\bar{r}_{*k-1})\Phi_s(j,k)-(\bar{r}_t-\bar{r}_{*k})\Phi_s(j,k-1)}{\Delta \bar{r}_*-C_s(\bar{a})(\bar{r}_t-\bar{r}_{*k-1})(\bar{r}_t-\bar{r}_{*k})}.
\end{equation}

\begin{figure}[htbp]
	\centering
	\subfigure[]{\label{fig:SOD-12-0}
		\begin{minipage}[t]{0.45\linewidth}
			\centering
			\includegraphics[width=1\linewidth]{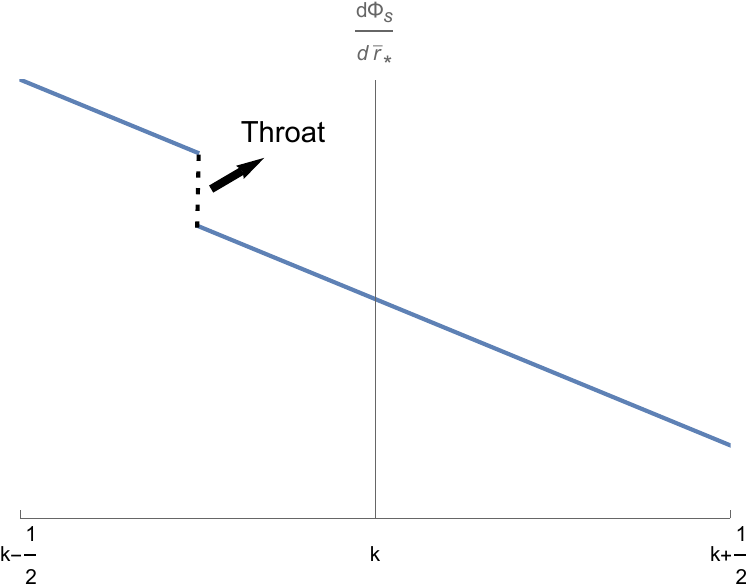}
		\end{minipage}
	}
	\subfigure[]{\label{fig:FOD-12-0}
		\begin{minipage}[t]{0.45\linewidth}
			\centering
			\includegraphics[width=1\linewidth]{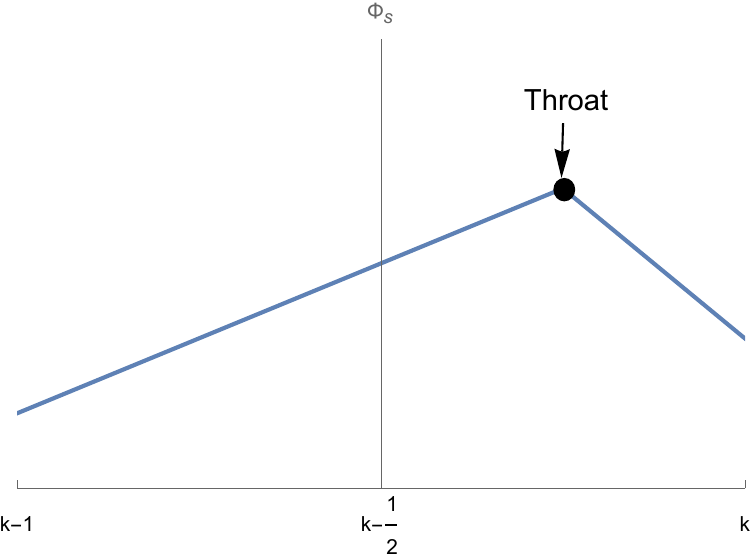}
		\end{minipage}
	}
	
	\caption{The schematic diagrams of $\dif\Phi_s/\dif\bar{r}_*$ and $\Phi_s$ varying with the tortoise coordinate within the grid points $(j,k-1)$ and $(j,k+1)$. These diagrams are plotted according to Eq.~\eqref{eq:discon}, and the wormhole throat $\bar{r}_t$ is within the range of $(\bar{r}_{*k-\frac{1}{2}},\bar{r}_{*k})$.}
	\label{fig:OD-1/2-0}
\end{figure}

When $\bar{r}_t\in(\bar{r}_{*k-\frac{1}{2}},\bar{r}_{*k})$, the discontinuity of $\dif\Phi_s/\dif\bar{r}_*$ at the throat directly affects the calculation of $\dif^2\Phi_s/d\bar{r}_*^2$ and $\dif\Phi_s(j,k-\frac{1}{2})/\dif\bar{r}_*$.
In Fig.~\ref{fig:OD-1/2-0}, we present the plots of $\Phi_s$ and $\dif\Phi_s/\dif\bar{r}_*$ varying with the tortoise coordinate within the corresponding grid points.
According to Fig.~\ref{fig:SOD-12-0} and Eq.~\eqref{eq:discon}, we obtain
\begin{equation}\label{eq:SOD-12-0}
    \frac{\dif^2\Phi_s(j,k)}{\dif \bar{r}_*^2}=\frac{\frac{\dif\Phi_s(j,k+\frac{1}{2})}{\dif\bar{r}_*}-\left[\frac{\dif\Phi_s(j,k-\frac{1}{2})}{\dif\bar{r}_*}+C_s(\bar{a})\Phi_s(j,\bar{r}_t)\right]}{\Delta\bar{r}_*}=\frac{\frac{\dif\Phi_s(j,k+\frac{1}{2})}{\dif\bar{r}_*}-\frac{\dif\Phi_s(j,k-\frac{1}{2})}{\dif\bar{r}_*}}{\Delta\bar{r}_*}-\frac{C_s(\bar{a})\Phi_s(j,\bar{r}_t)}{\Delta\bar{r}_*}.
\end{equation}
And by using Fig.~\ref{fig:FOD-12-0} and Eq.~\eqref{eq:discon}, we derive the equations,
\begin{equation}\label{eq:fod-12-0}
    \Phi_s(j,k)-\Phi_s(j,k-1)=\frac{\dif\Phi_s(j,k-\frac{1}{2})}{\dif\bar{r}_*}(\bar{r}_t-\bar{r}_{*k-1})+\frac{\dif\Phi_s(j,k)}{\dif\bar{r}_*}(\bar{r}_{*k}-\bar{r}_t)
\end{equation}
and
\begin{equation}
    \frac{\dif\Phi_s(j,k)}{\dif\bar{r}_*}=\frac{\dif\Phi_s(j,k-\frac{1}{2})}{\dif\bar{r}_*}+C_s(\bar{a})\Phi_s(j,\bar{r}_t).
\end{equation}
After simplifying these two equations, we get
\begin{equation}\label{eq:FOD-12-0}
    \frac{\dif\Phi_s(j,k-\frac{1}{2})}{\dif\bar{r}_*}=\frac{\Phi_s(j,k)-\Phi_s(j,k-1)}{\Delta \bar{r}_*}+C_s(\bar{a})\Phi_s(j,\bar{r}_t)\frac{\bar{r}_t-\bar{r}_{*k}}{\Delta\bar{r}_*}.
\end{equation}
Combining Eq.~\eqref{eq:SOD-12-0} and Eq.~\eqref{eq:FOD-12-0}, we obtain
\begin{equation}
\begin{split}
    \frac{\dif^2\Phi_s(j,k)}{\dif \bar{r}_*^2}
    &=\frac{\frac{\dif\Phi_s(j,k+\frac{1}{2})}{\dif\bar{r}_*}-\frac{\dif\Phi_s(j,k-\frac{1}{2})}{\dif\bar{r}_*}}{\Delta\bar{r}_*}\\
    &=\frac{\Phi_s(j,k+1)-2\Phi_s(j,k)+\Phi_s(j,k-1)}{\Delta\bar{r}_*^2}-C_s(\bar{a})\Phi_s(j,\bar{r}_t)\frac{\bar{r}_t-\bar{r}_{*k-1}}{\Delta\bar{r}_*^2}.\\
\end{split}
\end{equation}
whose form is consistent with Eq.~\eqref{eq:SOD-1-12}, and $\Phi_s(j,\bar{r}_t)$ is also consistent with Eq.~\eqref{eq:phisthroat-1-12}.

Therefore, when $\bar{r}_t \in (\bar{r}_{*k-1}, \bar{r}_{*k})$, Eq.~\eqref{eq:SOD-1-12} and Eq.~\eqref{eq:phisthroat-1-12} determine the difference form of $\dif^2\Phi_s/d\bar{r}_*^2$ at the grid point $(j,k)$.
Using the same calculation method, when $\bar{r}_t \in (\bar{r}_{*k}, \bar{r}_{*k+1})$, we obtain the difference form of $\dif^2\Phi_s/d\bar{r}_*^2$ at the grid point $(j,k)$ as follows:
\begin{equation}\label{eq:SOD+0+1}
    \frac{\dif^2\Phi_s(j,k)}{\dif \bar{r}_*^2}=\frac{\Phi_s(j,k+1)-2\Phi_s(j,k)+\Phi_s(j,k-1)}{\Delta\bar{r}_*^2}-C_s(\bar{a})\Phi_s(j,\bar{r}_t)\frac{\bar{r}_{*k+1}-\bar{r}_t}{\Delta\bar{r}_*^2},
\end{equation}
where
\begin{equation}\label{eq:phisthroat+0+1}
    \Phi_s(j,\bar{r}_t)=\frac{(\bar{r}_t-\bar{r}_{*k})\Phi_s(j,k+1)-(\bar{r}_t-\bar{r}_{*k+1})\Phi_s(j,k)}{\Delta \bar{r}_*-C_s(\bar{a})(\bar{r}_t-\bar{r}_{*k})(\bar{r}_t-\bar{r}_{*k+1})}.
\end{equation}

Finally, it yields the difference equation of Eq.~\eqref{eq:Peg-dl-B} substituting Eqs.~\eqref{eq:SOD-1-12},~\eqref{eq:phisthroat-1-12},~\eqref{eq:SOD+0+1} and ~\eqref{eq:phisthroat+0+1} into Eq.~\eqref{eq:Peg_FD-B}.
When $\bar{r}_t \in (\bar{r}_{*k-1}, \bar{r}_{*k})$, the difference equation at the grid point $(j,k)$ is
\begin{eqnarray}\label{eq:Peg_FD-dat-B}
& &
-\frac{\Phi_s(j+1,k)-2\Phi_s(j,k)+\Phi_s(j-1,k)}{\Delta \bar{t}^2}+\frac{\Phi_s(j,k+1)-2\Phi_s(j,k)+\Phi_s(j,k-1)}{\Delta \bar{r}_*^2}\nonumber \\
& &-\bar{V}_s(k)\Phi_s(j,k)+\frac{C_s(\bar{a})(\bar{r}_{*k-1}-\bar{r}_t)\Phi_s(j,\bar{r}_t)}{\Delta \bar{r}^2_*}=0,
\end{eqnarray}
and when $\bar{r}_t \in (\bar{r}_{*k}, \bar{r}_{*k+1})$, the difference equation is
\begin{eqnarray}\label{eq:Peg_FD-uat-B}
& &
-\frac{\Phi_s(j+1,k)-2\Phi_s(j,k)+\Phi_s(j-1,k)}{\Delta \bar{t}^2}+\frac{\Phi_s(j,k+1)-2\Phi_s(j,k)+\Phi_s(j,k-1)}{\Delta \bar{r}_*^2}\nonumber \\
& &-\bar{V}_s(k)\Phi_s(j,k)-\frac{C_s(\bar{a})(\bar{r}_{*k+1}-\bar{r}_t)\Phi_s(j,\bar{r}_t)}{\Delta \bar{r}^2_*}=0.
\end{eqnarray}

\end{appendices}



\bibliographystyle{unsrturl}
\bibliography{references}

\end{document}